\documentclass[10pt,journal]{IEEEtran}
\usepackage{amssymb}
\usepackage{dsfont}
\usepackage{amsfonts}
\usepackage{url}
\usepackage{eufrak}
\usepackage{amsfonts,subfigure,multicol,color,verbatim,
graphicx,cite,amssymb,amsmath,cases,bm,algorithm,
algorithmic,xcolor,multirow,array}
\usepackage{caption}
\usepackage{subfigure}
\usepackage{booktabs}
\usepackage{hyperref}
\usepackage{xcolor}

\usepackage{cite}

\allowdisplaybreaks
\begin{document}

\title{Transmission Delay Minimization for NOMA-Based F-RANs}
\author{Yuan~Ai, Xidong Mu, \emph{Member,~IEEE}, Pengbo~Si, \emph{Senior Member,~IEEE}, Yuanwei~Liu, \emph{Fellow,~IEEE}
\thanks{Y. Ai, P. Si are with the School of Information Science and Technology, Beijing University of Technology, Beijing 100124, China (e-mail: aiyuan@bjut.edu.cn; sipengbo@bjut.edu.cn).	\emph{(Corresponding author: Pengbo Si.)}

X. Mu is with the Centre for Wireless Innovation (CWI), Queen’s University Belfast, Belfast, BT3 9DT, U.K. (e-mail: x.mu@qub.ac.uk).

Y. Liu is with the Department of Electrical and Electronic Engineering, The
University of Hong Kong, Hong Kong (e-mail: yuanwei@hku.hk).
}
%\thanks{This work was supported in part by the National Natural Science Foundation of China (Grant No.61361166005) and the National Basic Research Program of China (973 Program) (Grant No. 2013CB336600).}
}

\maketitle

\begin{abstract}
A novel non-orthogonal multiple access (NOMA) based low-delay service framework is proposed for fog radio access networks (F-RANs). Fog access points (FAPs) leverage NOMA for local delivery of cached content, while the cloud access point employs NOMA to simultaneously push content to FAPs and directly serve users. Based on this model, a delay minimization problem is formulated by jointly optimizing user association, cache placement, and power allocation. To address this non-convex mixed-integer nonlinear programming problem, an alternating optimization (AO) algorithm is developed, which decomposes the original problem into two subproblems, namely joint user association and cache placement, and power allocation. In particular, a low-complexity algorithm is designed to optimizing the user association and cache placement strategy using the McCormick envelope theory and Lagrangian partial relaxation. The power allocation is optimized by invoking the successive convex approximation. Simulation results reveal that: 1) the proposed AO-based algorithm effectively balances between the achieved performance and computational efficiency, and 2) the proposed NOMA-based F-RANs framework significantly outperforms orthogonal multiple access-based F-RANs systems in terms of average transmission delay in different scenarios.
\end{abstract}


\section{Introduction}
The rapid expansion of data demand, coupled with the emergence of bandwidth-intensive applications, is driving the development of the next generation of moible communication networks. This evolution presents substantial challenges, including the management of heterogeneous service data traffic, accommodating ultra-massive connectivity, and achieving ultra-high throughput alongside minimal latency \cite{mu2023noma}. To meet these challenges, the development of advanced multiple access technology, collectively termed next generation multiple access (NGMA), is anticipated. These schemes are expected to support a vast number of users more efficiently in terms of both resource utilization and computational complexity compared to current multiple access technologies \cite{liu2022evolution}. Among these advanced schemes, power-domain non-orthogonal multiple access (NOMA) has emerged as a crucial technology, garnering significant attention from both academia and industry \cite{ding2017survey}. By efficiently assigning multiple user signals to the same resource block via superposition coding, NOMA offers considerable performance enhancements over orthogonal multiple access (OMA) methods, particularly in terms of network capacity, user connectivity, and latency reduction \cite{mu2021energy}.

On the other hand, fog radio access networks (F-RANs) have emerged as a critical technology and architectural innovation in 6G, offering solutions for the diverse demands of heterogeneous devices through advanced and efficient resource management strategies \cite{kaneko2020opportunities}. F-RANs are designed to address the latency and signaling overhead challenges inherent in cloud radio access networks (C-RANs) by partially shifting network intelligence, such as computing and storage capabilities, closer to edge devices \cite{ai2018edge}. This is achieved through the deployment of fog access points (FAPs) that combine distributed caching and advanced signal processing functionalities, enabling collaborative cloud-edge processing. To further enhance these benefits, integrating power-domain NOMA technology within F-RANs is expected to meet the extraordinary requirements for high access speeds and low latency. NOMA can be incorporated into F-RANs to facilitate content delivery from FAPs to mobile users, as well as to manage task offloading when computational tasks are requested by users. As a result, the fusion of NOMA with F-RANs has sparked a new research area known as NOMA-based F-RANs, which holds significant promise for enhancing spectral efficiency and reducing transmission delays in next-generation wireless systems \cite{zhang2018resource}.
\subsection{Related Works}
\subsubsection{Studies on NOMA-based F-RANs} 
Recent research extensively investigated  the potential benefits of integrating NOMA to improve the performance of F-RANs. In \cite{yang2022new}, a detailed analysis of the network architecture and core modules of NOMA F-RANs was provided, with a focus on the application of artificial intelligence (AI)-enabled methods to optimize resource allocation through latent feature extraction and cooperative caching. Similarly, Guo \emph{et al.} \cite{guo2022cross} investigated the benefits of integrating cache-aided multicast transmissions with NOMA in F-RANs, using index coding to minimize transmission energy. Furthermore, Ai \emph{et al.} \cite{ai2023joint} introduced a cost-efficent resource allocation framework specifically designed for NOMA-based F-RANs, achieving a significant trade-off between network throughput and computational complexity. 

\subsubsection{Studies on Low-Delay Optimization}

Minimizing transmission delay remains a critical challenge in the design of F-RANs, particularly when integrating NOMA. Recent advancements have explored the synergy of NOMA with caching and computing to address these multi-faceted challenges. Shen \emph{et al.} \cite{shen2023power} proposed a comprehensive framework integrating task offloading and caching in NOMA-assisted networks, utilizing deep reinforcement learning to optimize resource allocation and reduce latency. Moreover, Ye \emph{et al.} \cite{ye2023intelligent} introduced a hierarchical network slicing approach for 6G systems, integrating NOMA and multi-dimensional resource allocation to enhance efficiency and meet the stringent requirements of latency-critical applications. Yin \emph{et al.} \cite{yin2023joint} devised a joint optimization strategy for user pairing, power allocation, and content server placement in NOMA-assisted wireless caching networks, improving hit probabilities under dynamic channel conditions. Furthermore, Dong \emph{et al.} \cite{dong2023joint} focused on reliability-aware services in NOMA-based networks, incorporating power allocation and task offloading to ensure ultra-reliable low-latency communications. Lim \emph{et al.} \cite{lim2023joint} tackled the challenge of imperfect successive interference cancellation (SIC) in mmWave-NOMA, proposing a cross-entropy-based clustering and beamforming algorithm to optimize system performance. Qin \emph{et al.}~\cite{qin2024collaborative} proposed a cluster-NOMA framework for space-air-ground edge computing, utilizing multi-agent learning to optimize task offloading, caching, and routing for reduced IoT latency. Fang \emph{et al.}~\cite{fang2024joint} developed a joint offloading and caching scheme in multi-cell NOMA networks, employing deep reinforcement learning and clustering to minimize content delivery delay. Dou \emph{et al.}~\cite{dou2024integrated} introduced a NOMA-assisted system for integrated sensing and offloading, optimizing beamforming and resource allocation to minimize energy while preserving sensing performance.

Despite these contributions, existing research often addresses isolated optimization dimensions, such as caching or task offloading, without fully exploiting their interplay, leading to suboptimal solutions or high computational complexity. In contrast, our proposed framework comprehensively integrates user association, cache placement, and power allocation in the NOMA-enabled F-RANs. By employing NOMA at both FAPs and the cloud access point (CP), we enhance cloud-fog collaboration to achieve low-delay services. Different from all existing works that apply NOMA exclusively at the fog/edge tier, the proposed framework innovatively introduces power-domain NOMA at the CP to realize a dual-purpose `push-and-deliver' transmission strategy. This enables the CP to simultaneously push uncached popular files to multiple FAPs and deliver requested content directly to UEs using a single superimposed signal---drastically reducing fronthaul latency compared with orthogonal schemes that require sequential transmissions. Table \ref{table:contributions} compares our work with these recent studies, highlighting our unique contributions in jointly optimizing multiple dimensions and leveraging NOMA across both edge and cloud tiers for superior latency performance.

\begin{table*}[htbp]
		\vspace{-0.2cm}\tiny
		\renewcommand{\arraystretch}{0.8}
		\caption{Comparison of Our Contributions with the State-of-the-Art Works}
		\centering
		\resizebox{\textwidth}{!}{
		\begin{tabular}{lcccccccccc}
		\toprule
		\textbf{Aspect} & \textbf{\cite{shen2023power}} & \textbf{\cite{ye2023intelligent}} & \textbf{\cite{yin2023joint}} & \textbf{\cite{dong2023joint}} & \textbf{\cite{lim2023joint}} & \textbf{\cite{qin2024collaborative}} & \textbf{\cite{fang2024joint}} & \textbf{\cite{dou2024integrated}} & \textbf{Proposed} \\
		\midrule
		User Association & $\times$ & $\times$ & $\checkmark$ & $\times$ & $\checkmark$ & $\times$ & $\checkmark$ & $\times$ & $\checkmark$ \\
		Cache Placement & $\times$ & $\checkmark$ & $\checkmark$ & $\checkmark$ & $\times$ & $\checkmark$ & $\times$ & $\times$ & $\checkmark$ \\
		Power Allocation & $\checkmark$ & $\times$ & $\checkmark$ & $\checkmark$ & $\checkmark$ & $\checkmark$ & $\checkmark$ & $\checkmark$ & $\checkmark$ \\
		NOMA in FAP & $\checkmark$ & $\checkmark$ & $\checkmark$ & $\checkmark$ & $\checkmark$ & $\checkmark$ & $\checkmark$ & $\checkmark$ & $\checkmark$ \\
		NOMA in CP & $\times$ & $\times$ & $\times$ & $\times$ & $\times$ & $\times$ & $\times$ & $\times$ & $\checkmark$ \\
		Low Delay Service & $\times$ & $\checkmark$ & $\times$ & $\times$ & $\times$ & $\checkmark$ & $\checkmark$ & $\times$ & $\checkmark$ \\
		Cloud-Fog Collaboration & $\times$ & $\checkmark$ & $\times$ & $\times$ & $\times$ & $\checkmark$ & $\checkmark$ & $\checkmark$ & $\checkmark$ \\
		Joint Assoc./Caching/Power Alloc. & $\times$ & $\times$ & $\times$ & $\times$ & $\times$ & $\times$ & $\times$ & $\times$ & $\checkmark$ \\
		\bottomrule
		\end{tabular}
		}
		\label{table:contributions}
		\vspace{-0.2cm}
\end{table*}

\subsection{Motivations and Contributions}

While NOMA offers potential benefits in terms of efficient resource allocation, the integration of NOMA with F-RANs introduces new challenges that must be addressed to fully realize these benefits. The tightly coupled nature of computation and communication resources in NOMA-based F-RANs complicates the management of these resources. Variability in user demand, the limited capacity of FAPs, and the coexistence of diverse user access modes all complicate the resource allocation process. These challenges are further exacerbated by the need for real-time decision-making, which is essential for meeting the stringent low-delay requirements of modern networks. As a result, joint user association and resource allocation strategies become critical, as they must dynamically account for the varying nature of both communication and computation resources to minimize transmission delay and optimize network performance.

Despite extensive research on NOMA and F-RANs individually, and some exploration of their integration, there remains a critical gap in addressing these combined challenges from a delay perspective. Existing studies on NOMA-based F-RANs often overlook the complexities introduced by the need for joint resource management, particularly the non-convex optimization problem that arises when designing user association, cache placement, and power allocation strategies under practical constraints such as FAP caching capacity and power budgets. Traditional approaches often fail to adequately address these issues or result in solutions that are computationally prohibitive, limiting their practical applicability.

Given the challenges in managing resource allocation and reducing delays in F-RANs, this paper introduces a novel NOMA-based F-RAN framework. This framework leverages NOMA's flexible resource management capabilities within a fog-cloud architecture. A joint resource allocation problem is formulated that integrates user association, cache management, and power allocation, addressing these interconnected challenges through advanced optimization techniques. The solution balances the need for performance enhancement with computational efficiency, as demonstrated by the significant reduction in transmission delay observed in our simulations. The main contributions are as follows:
\begin{enumerate}
\item We propose a novel low-delay service framework for NOMA-based F-RANs, where FAPs leverage NOMA for local delivery of cached content, while the CP employs NOMA to simultaneously push content to FAPs and directly serve users. This framework explicitly accounts for the constraints of FAP caching capacity and power allocation. We formulate a non-convex mixed-integer nonlinear programming (MINLP) problem aimed at minimizing the average delay through joint optimization of user association, cache placement, and power allocation.
\item Given the non-deterministic polynomial-time (NP)-hard nature of the problem, an alternating  optimization (AO) algorithm is developed, which decomposes the original problem into two subproblems, namely a) \emph{Joint user association and cache placement} subproblem, and b) \emph{Power allocation} optimization subproblem. For the joint user association and cache placement, we introduce a low-complexity algorithm based on the \emph{McCormick envelope theory} and \emph{Lagrangian partial relaxation}, effectively addressing the multidimensional resource constraints. For the non-convex power allocation problem, we propose a \emph{successive convex approximation (SCA)}-based algorithm, thereby significantly reducing the computational complexity while maintaining near-optimal performance.
\item We demonstrate that the proposed NOMA-based F-RANs significantly outperforms OMA-based F-RANs systems in terms of average transmission delay in different scenarios. Additionally, the proposed AO algorithm strikes an effective balance between the achieved system performance and computational efficiency, making it a viable candidate for practical deployment in future wireless networks.
\end{enumerate}

\subsection{Organization}
The structure of this paper is organized as follows: Section \ref{system} presents the system model for NOMA-based F-RANs and formulates the problem addressed in this study. In Section \ref{AO}, we develop a low-complexity algorithm that integrates the subproblems of user association, cache placement, and power allocation into a cohesive framework. Section \ref{VI} provides numerical results and assesses the performance of the proposed algorithm. Finally, Section \ref{VII} concludes the paper with a summary of the findings.

\section{System Model}\label{system}
\subsection{System Description}

\begin{figure}[t]
	\centering
	\includegraphics[width=0.4\textwidth]{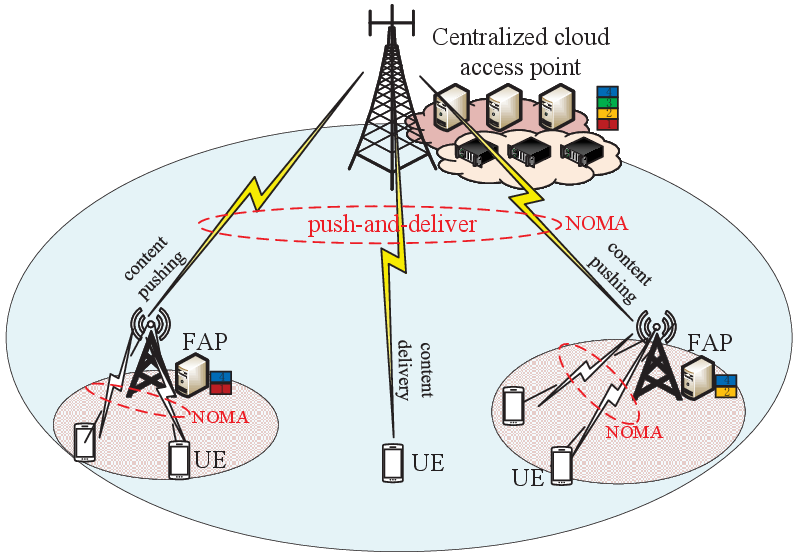}\\
	\caption{An illustration of the NOMA-based F-RANs.}\label{Architecture}
   \end{figure}

As depicted in Fig. \ref{Architecture}, we examine a downlink NOMA-based F-RANs configuration consisting of a centralized CP, $N$ FAPs, and $K$ user equipments (UEs).\footnote{This architecture is particularly suited for practical scenarios such as IoT networks, smart cities, and vehicular networks, where low-latency content delivery and efficient spectrum utilization are critical. For instance, in IoT networks, FAPs serve numerous devices with cached content, while in vehicular networks, roadside units acting as FAPs provide localized processing in high-mobility settings.} The CP and FAPs are situated at predetermined positions within a circular region $D$ with radius $R_D$, while the UEs are distributed uniformly throughout this area. Each device is assumed to be equipped with a single antenna. Meanwhile, the collection of FAPs and the CP is represented by the set $\mathcal{N}=\{0, 1, 2, \ldots, N\}$, where the element 0 in $\mathcal{N}$ specifically denotes the CP. The set of FAPs (excluding the CP) is denoted by $\mathcal{N}^+ = \mathcal{N} \setminus \{0\}$. The set of UEs is denoted by $\mathcal{K}=\{1, 2, 3, \ldots, K\}$. Each FAP has caching capabilities and can deliver data to UEs by utilizing cached content or through edge processing. Multiple UEs are simultaneously served by each FAP using the NOMA protocol. The FAPs are connected to the CP via a shared wireless fronthaul link. The NOMA protocol allows the CP to execute the push-and-delivery strategy \cite{ding2018noma}, enhancing spectrum efficiency in wireless caching. This strategy enables concurrent wireless fronthaul transmissions between the CP and FAPs and wireless access transmissions from the CP to UEs. If the requested content is not available at the local FAP, the UE is served directly by the CP. NOMA is applied to achieve dual objectives: content pushing to FAPs for future use and immediate content delivery to UEs by the CP. The CP and FAPs operate on separate frequency bands to eliminate cross-tier interference. The CP uses its dedicated band for dual-purpose NOMA transmission, while FAPs share a distinct band for NOMA-based access. This design decouples interference while maximizing intra-tier spectral efficiency via NOMA.

\subsection{Cache Model}
Consider a set of popular content files denoted by $\mathcal{F}=\{1, 2, 3, \ldots, F\}$, which are distributed at the CP according to the Zipf distribution. Each requested content file $f$ has a packet size of $s_f$. The CP can access the entire content library, while each FAP is capable of prefetching and storing selected content files from the CP's library in its local memory. When serving users, if the requested content is not stored locally, the FAPs can directly download the data from the CP. Additionally, FAPs can decode the data from the received signals and deliver the requested content to users, thereby fulfilling their demands. Each FAP is equipped with a cache of capacity $S_n$. To optimize content delivery, each FAP $n$ initially downloads the most popular content files $f\in\mathcal{F}$ from the CP and stores them within its cache according to its capacity. The network controller determines which content should be cached at each FAP $n$. The caching decision is represented as the binary decision variable $c_{n,f}\in \{0,1\}$, which depicts whether the content $f$ should be cached or not at FAP $n$, where
\begin{equation}\label{cnf}
	c_{n,f}\in \{ 0, 1\}, f\in \mathcal{F}, n\in  \mathcal{N}^+.
\end{equation}
In this context, $c_{n,f}=1$ indicates that content $f$ is stored at FAP $n$, while $c_{n,f}=0$ indicates that it is not. If content $f$ is already cached at FAP $n$, there is no need for FAP $n$ to retrieve it from the CP when it is requested by associated UEs. However, if $c_{n,f}=0$, FAP $n$ must fetch the content from the CP upon receiving a request from its associated UEs. The cache size constraint for each FAP $n$ is expressed as 
\begin{equation}\label{Snf} 
	\sum_{f\in\mathcal{F}} c_{n,f} s_f \leq S_n, \quad n\in \mathcal{N}^+. 
\end{equation} 
The caching decision for all FAPs is represented by the vector $\mathbf{c}\triangleq (c_{n,f})_{f\in\mathcal{F},n\in \mathcal{N}}$.

\subsection{NOMA-based Communication Model}
\subsubsection{Transmission from FAP to UE}
For $n\in\mathcal{N}$, $k\in \mathcal{K}$, let $x_{n,k}$ denote whether UE choosing to associated with node $n$, where
\begin{equation}\label{xnk}
	x_{n,k}\in \{ 0, 1\}, n\in  \mathcal{N},k\in \mathcal{K}.
\end{equation}
Here, $x_{n,k}=1$ means that the $k$-th UE is served by the access node $n$, and otherwise if $x_{n,k}=0$. We denote $\mathbf{x}\triangleq (x_{n,k})_{n\in\mathcal{N},k\in  \mathcal{K}}$ as the UE association vector. It is assumed that the individual UEs can be associated with one access node at the same time. The UEs associated with the same node are considered as a cluster. Let $\mathcal{K}_n$ represent as the cluster formed by node $n$. In any cluster $\mathcal{K}_n$ that contains $t$ UEs, where $t=\left|\mathcal{K}_n\right|$, the received signal by UE $k$ associated with FAP $n$ is given by
\begin{equation}\label{ynk}
	\begin{split}
	y_{n,k}=&h_{n,k}\sqrt {p_{n,k}}s_{n,k}+
	\underbrace{h_{n,k}\sum_{i=1,i \ne k}^{t}\sqrt {p_{n,i}}s_{n,i}}_{\text{intra-FAP interference}}\\
	&+\underbrace{\sum_{j = 1,j \ne n}^N {\sum_{i = 1}^t {{h_{j,k}} \sqrt {{p_{j,i}}} {s_{j,i}}} }}_{\text{inter-FAP interference}}+n_{n,k},
	\end{split}
\end{equation}
where $h_{n,k}\in \mathbb{C}$ represents the complex channel gain between UE $k$ and FAP $n$, $p_{n,k}\geq 0$ denotes the power allocation for user $k$ by FAP \(n\), \(s_{n,k}\) is the transmitted signal for UE \(k\), normalized such that \(\mathbb{E}[|s_{n,k}|^2] = 1\), and \(n_{n,k} \sim \mathcal{CN}(0, \sigma^2)\) is the additive white Gaussian noise (AWGN) at UE \(k\), with variance \(\sigma^2 > 0\). In addition, the channel coefficient for UE $k$ in cluster $\mathcal{K}_n$, where $t = \left|\mathcal{K}_n\right|$, is given by ${{\left| {h_{n,k}} \right|}^2} = {{\left| {g_{n,k}} \right|}^2} L_{n,k}^{-1}$, where $g_{n,k} \in \mathbb{C}$ denotes the small-scale fading, with \(|g_{n,k}|^2 \sim \text{Exp}(1)\) for Rayleigh fading, $L_{n,k}$ represents the large-scale path loss coefficient. In addition, the power coefficients within any cluster $\mathcal{K}_n$ are constrained by $\sum_{i=1}^{t} p_{n,i} = p_{n}$ and $p_{n,i} \geq 0$ for all $i \in \mathcal{K}_n$.

The inter-FAP interference \(I_{n,k}\) is defined as:
\begin{equation}\label{interinf}
	I_{n,k} = \sum_{j = 1,j \ne n}^N {\sum_{i = 1}^t {{{\left| {{h_{j,k}}} \right|}^2}  {{p_{j,i}}} } }=\sum_{j = 1,j \ne n}^N {{\left| {{h_{j,k}}} \right|}^2} p_j,
\end{equation}
where \(p_j = \sum_{i \in \mathcal{K}_j} p_{j,i}\) is the total transmit power of FAP \(j\).

In the proposed NOMA-based F-RANs, intra-FAP interference results from co-channel transmissions under the NOMA scheme, and SIC is employed to mitigate this interference \cite{zhao2017spectrum}.  According to the principles of SIC and NOMA, the UE with strong channel conditions cancel out interference from UEs with weaker conditions, while UEs with weaker channels must consider the signals from stronger UEs as additional noise. Consequently, the signal-to-interference-plus-noise ratio (SINR) for UE $k$ associated with FAP $n$ is given by
\begin{eqnarray}\label{sinrnk}
	{\gamma_{n,k}}=\frac{ {{\left| {{h_{n,k}}} \right|}^2}{p_{n,k}} }
	{ \sum\limits_{i\in \mathcal{K}_n \backslash\{k\}: \pi(i)>\pi(k)}{{\left| {{h_{n,k}}} \right|}^2}{p_{n,i}}+I_{n,k}+\sigma^2}.
\end{eqnarray}
where \( \pi(k) \in \{1, 2, \ldots, t\} \) denotes the decoding order position of UE \( k \), with higher \( \pi(k) \) indicating a stronger channel condition.

Note that the conventional decoding order in single-cell NOMA systems, which relies solely on the order of channel gains, cannot be directly applied in the proposed multi-FAP scenario. In particular, both intra-FAP interference and inter-FAP interference must be accounted for in the effective channel conditions. As discussed in \cite{you2020note}, one of the challenges in establishing the SIC decoding order lies in the fact that interference in a given FAP depends on the power and time-frequency resources allocated not only locally but also in other FAPs. Hence, the exact decoding order is generally not known \textit{a priori}, because it is inherently coupled with the interference levels arising from resource usage decisions across multiple FAPs.

Despite these complications, the fundamental principle for SIC decoding in NOMA still applies. Formally, consider two UEs, $k$ and $m$, within the same cluster $\mathcal{K}_n$. Suppose UE $k$ has a better channel condition than UE $m$, denoted by $m < k$. Perfect SIC occurs if the decoding rate of UE $k$ for UE $m$'s signal meets or exceeds the rate at which UE $m$ decodes its own signal, that is, ${R_{n,k \to m}} \ge {R_{n,m \to m}}$. In our NOMA-based F-RAN setup, each FAP applies SIC on a per-cluster basis, while treating the aggregated inter-FAP interference as noise, similar to the single-cell case. Building on this principle from \cite{wang2019user}, we adopt the following optimal decoding order for cluster $\mathcal{K}_n$. Then, the decoding sequence can be expressed as
\begin{equation}\label{condition}
	\mathcal{Q}(\mathcal{K}_n)\triangleq \frac{ I_{n,1}+\sigma^2}{ {\left| {{h_{n,1}}} \right|}^2} \geq \frac{ I_{n,2}+\sigma^2}{ {\left| {{h_{n,2}}} \right|}^2} \geq...\geq \frac{ I_{n,t}+\sigma^2}{ {\left| {{h_{n,t}}} \right|}^2}.
\end{equation} 

Based on \eqref{sinrnk}, the transmission rate of UE $k$ at FAP $n$ can be expressed as
\begin{equation}\label{raternk}
		R_{n,k}=B \log_2 (1+{\gamma_{n,k}}),
\end{equation}
where \(B > 0\) is the total channel bandwidth in hertz (Hz).

\subsubsection{Transmission from CP to UE}
If content requirements of the UE cannot be met by the FAP mode, the system will utilize the CP mode to process the request of UE, i.e., $x_{0,k}=1$. In this scenario, the CP employs NOMA technology to execute the push and deliver strategy, ensuring efficient content distribution. Specifically, the CP transmits a superimposed signal comprising the following two components: 1) Direct content delivery: Files directly requested by UEs that cannot be served by FAPs; 2) Content pushing to FAPs: Files that are not cached at the FAPs but are requested by UEs associated with those FAPs. This simultaneous transmission on the same resource block is enabled exclusively by NOMA’s superposition coding. Under OMA, these operations would require two orthogonal time slots, significantly increasing fronthaul delay. We define the set of files that need to be pushed by the CP as \(\mathcal{F}^{\text{push}}\). A file \(f \in \mathcal{F}\) is included in \(\mathcal{F}^{\text{push}}\) if it satisfies the following condition: 
\begin{equation}\label{pushCP}
    \mathcal{F}^{\text{push}} = \left\{ f \in \mathcal{F} \ \big| \ \sum_{n \in \mathcal{N}} \sum_{k \in \mathcal{K}} x_{n,k} r_{k,f} (1 - c_{n,f}) \geq 1 \right\},
\end{equation}
where \(f \in \mathcal{F}\) represents a file in the content library, \(x_{n,k} = 1\) indicates that UE \(k\) is associated with FAP \(n\), \(r_{k,f} = 1\) indicates that UE \(k\) requests file \(f\), \(c_{n,f} = 0\) indicates that file \(f\) is not cached at FAP \(n\). Thus, \(\mathcal{F}^{\text{push}}\) contains all files that need to be pushed by the CP to the FAPs.

Given the decoding complexity associated with NOMA technology, this paper assumes a simplified scenario where the CP serves only one UE for direct content delivery while simultaneously pushing the required content to FAPs. In this setup, the received signal by UE $k$ associated with the CP directly is given by 
\begin{equation}\label{ynk00}
	y_{0,k}=h_{0,k}\sqrt {p_{0,k}} s_{0,k} +\sum_{f \in \mathcal{F}^{\text{push}}} \sqrt {p_{0,f}} h_{0,k}\overline{s}_{0,f} +n_{0,k},
\end{equation}
where $h_{0,k} \in \mathbb{C} $ is the complex channel gain between the CP and UE $k$, with ${{\left| {{h_{0,k}}} \right|}^2}={{\left| {{g_{0,k}}} \right|}^2}L^{-1}_{0,k}$, similar to $h_{n,k}$. $s_{0,k}$ is the transmitted signal for the $k$-th user, $\overline{s}_{0,f} $ denotes the signal that represents the information contained in file $f$, normalized such that \(\mathbb{E}[|s_{0,k}|^2] = 1\) and \(\mathbb{E}[|\overline{s}_{0,f}|^2] = 1\). \(p_{0,k}\) is the power allocated to UE \(k\) by the CP, \(p_{0,f}\) is the power allocated to file \(f \in \mathcal{F}^{\text{push}}\). For the CP, power allocation coefficients satisfy $p_{0,k}+\sum_{f \in \mathcal{F}^{\text{push}}} p_{0,f}\leq p_{0}$, where \(p_0 > 0\) is the total transmit power of the CP, ${p_{0,k}} \geq 0$ and ${p_{0,f}} \geq 0, \forall f \in \mathcal{F}$. \(n_{0,k} \sim \mathcal{CN}(0, \sigma^2)\) is the AWGN at UE \(k\), with variance \(\sigma^2 > 0\).

The signal intended for pushing files to the FAP is treated as interference by user $k$, and the received SINR of the UE $k$ directly served by the CP can be calculated as
\begin{equation}\label{sinr0k}
	{\gamma_{0,k}}=\frac{ {{\left| {{h_{0,k}}} \right|}^2}{p_{0,k}} }
	{ {{\left| {{h_{0,k}}} \right|}^2} \sum\limits_{f \in \mathcal{F}^{\text{push}}}{p_{0,f}}+\sigma^2}.
\end{equation}
Based on \eqref{sinr0k}, the transmission rate of UE $k$ at the CP can be expressed as
\begin{equation}\label{rater0k}
	R_{0,k}=B \log_2 (1+{\gamma_{0,k}}).
\end{equation}

\subsubsection{Transmission from CP to FAP}
Each FAP $n$ is capable of decoding additional files $f$ that are pushed from the CP through the superimposed coded signals. The decoding order using SIC is based on the popularity of the files, where a file $j$ with higher popularity (i.e., $j < f$) is decoded before a file $f$ with lower popularity. Assuming that all files with higher popularity, denoted as $j$ where $j < f$, have been successfully decoded and their effects have been subtracted, the SINR for FAP $n$ when decoding file $f$ is given by 
\begin{equation}\label{sinr0f}
	\gamma_{0,n}^f = \frac{ |h_{0,n}|^2 p_{0,f} }{ |h_{0,n}|^2 \sum_{i=f+1}^{|\mathcal{F}^{\text{push}}|} p_{0,i} + \sigma^2 },
\end{equation}
where \( |h_{0,n}|^2 = L(d_{0,n})^{-1} \), and \( L(d_{0,n}) \) is the path loss function. In the channel model between the CP and the FAP, small-scale multipath fading is not considered because the FAP and the CP can be connected via a line-of-sight (LoS) link, where large-scale path loss dominates. The corresponding downlink data rate is given by 
\begin{equation}\label{rater0f}
	R_{0,n}^f=B \log_2 (1+{\gamma_{0,n}^f}).
\end{equation}
If $R_{0,n}^f \geq R_f$, then file $f$ can be decoded and subtracted correctly at FAP $n$, where $R_f$ denotes the target data rate of file $f$. 

\subsection{Delay Model}

When UE \( k \) selects the edge cloud mode to access FAP \( n \), the transmission delay depends on whether the requested content is cached at the FAP. If the content is cached, the UE retrieves it directly from the FAP’s local cache without incurring additional delay via the fronthaul link. Otherwise, the FAP must fetch the content from the CP, introducing fronthaul delay. 
However, practical F-RAN systems incur additional latencies beyond pure transmission time, particularly due to the requisite control signaling and protocol interactions for link establishment \cite{peng2016fog}. To account for these practical constraints while maintaining the tractability of the system model, we introduce a constant overhead term $D_{\text{ov}}$, which represents the aggregated latency attributable to backhaul signaling and connection setup overhead\cite{zhao2019computation}.

To formulate the delay model, we define the request indicator variable \( r_{k,f} \in \{0,1\} \), where \( r_{k,f} = 1 \) if UE \( k \) requests content \( f \), and 0 otherwise, with the constraint \( \sum_{f=1}^F r_{k,f} = 1 \) ensuring each UE requests exactly one content file. Following the standard transmission models in \cite{dai2025joint}, the total delay from FAP \( n \) to UE \( k \), denoted \( D_{n,k} \), is given by:
\begin{equation}\label{Dnk}
\begin{split}
D_{n,k} &= D_{n,k}^F + (1 - c_{n,f}) D_{n,k}^B + D_{\text{ov}}\\
        &= \sum_{f=1}^F \left( \frac{r_{k,f} s_f}{R_{n,k}} + \frac{r_{k,f} (1 - c_{n,f}) s_f}{R_{0,n}^f} \right) + D_{\text{ov}},
\end{split}
\end{equation}
where:
\begin{itemize}
    \item \( D_{n,k}^F = \sum_{f=1}^F \frac{r_{k,f} s_f}{R_{n,k}} \): The access delay, representing the time to transmit the requested content from FAP \( n \) to UE \( k \).
    \item \( D_{n,k}^B = \sum_{f=1}^F \frac{r_{k,f} s_f}{R_{0,n}^f} \): The fronthaul delay, representing the time to fetch content \( f \) from the CP to FAP \( n \), incurred only when \( c_{n,f} = 0 \).
    \item \( D_{\text{ov}} \): The constant signaling and protocol overhead delay.
\end{itemize}

\textit{Remark 1:} In this study, we focus on minimizing the average transmission delay, which consists of the access delay and the wireless fronthaul delay. Following the established conventions in physical-layer resource management for F-RANs, queuing delay and SIC processing delay are omitted based on the following justifications: (i) Queuing Delay: We adopt a snapshot-based optimization approach focusing on the delivery of requested content files under a saturated buffer or a scheduled transmission period. In such a deterministic framework, queuing delay (which depends on stochastic arrival processes) is typically secondary to the transmission time when handling large content files. (ii) SIC Processing Delay: In practical NOMA systems with limited cluster sizes (e.g., 2 or 3 users), the iterative decoding latency is in the order of microseconds, which is negligible compared to the transmission and signaling overhead $D_{\text{ov}}$. Consequently, it is absorbed into the constant term $D_{\text{ov}}$ without loss of generality.

Despite the increase of the number of FAPs available for content service, the edge cloud remains unable to meet all user demands due to the limited resources of FAPs. Consequently, a significant amount of content must still be delivered through the centralized CP. The delay from the centralized CP to UE $k$ can be expressed as
\begin{eqnarray}\label{D0k}
	D_{0,k}=  \sum\limits_{f = 1}^F \frac{r_{k,f}  s_f }{R_{0,k}} + D_{\text{ov}}.
\end{eqnarray}
\subsection{Problem Formulation}
For the sake of improving the performance of the proposed NOMA based F-RANs, an optimization problem is formulated for minimizing the average system delay. By optimizing user association, cache placement, and power allocation, the problem can be formulated as follows:
\begin{subequations}\label{4_ProblemFormulation1}
	\begin{align}
		\mathop {\min }\limits_{\{ \bf{x},\bf{c},\bf{z} \} } &~~\displaystyle{ \frac{1}{K}\sum\limits_{n = 0}^N \sum\limits_{k = 1}^K   { x_{n,k} D_{n,k} } }\label{4_C0}\\ 	
		\text{s.t.}&~~\mathcal{Q}(\mathcal{K}_n),\forall k,\forall n,\label{C01order}\\
		      &~~\sum_{f=1}^{F} c_{n,f} s_f \leq S_n ,\forall n \in\mathcal{N}^+ ,\label{4_C01}\\
		      &~~\sum_{n\in\mathcal{N}} x_{n,k}=1,\forall k \in\mathcal{K},\label{4_C02}\\
		      &~~\sum_{k\in\mathcal{K}} x_{n,k}\leq F_n,\forall n \in\mathcal{N},\label{4_C03}\\
		      &~~c_{n,f}\in\{0,1\},\forall n \in\mathcal{N}^+,\forall f\in\mathcal{F},\label{4_C04}\\
		      &~~x_{n,k}\in\{0,1\},\forall n \in\mathcal{N},\forall k \in\mathcal{K},\label{4_C05}\\
		      &~~\sum\limits_{k = 1}^K  {p_{n,k}} \leq p_{n}^{max},\forall n \in\mathcal{N}^+,\label{4_C06}\\
			  &~~{p_{n,k}} \geq 0, \forall n \in\mathcal{N},\forall k \in\mathcal{K},\label{4_C07}\\
		      &~~p_{0,k}+\sum_{{f \in \mathcal{F}^{\text{push}}}}p_{0,f}\leq p_{0}^{max}, \label{4_C08}\\
			  &~~{p_{0,k}} \geq 0,{p_{0,f}} \geq 0, \forall k \in\mathcal{K}, \forall f \in \mathcal{F}.\label{4_C09}
	\end{align}
\end{subequations}
Constraint \eqref{C01order} ensures the optimal decoding order within each FAP's user cluster, guaranteeing that any UE can successfully decode the signal of a UE with weaker channel conditions. Constraint \eqref{4_C01} imposes the cache capacity limit of the FAP. Constraint \eqref{4_C02} represents the user association constraint, stipulating that each user can associate with only one FAP. Constraint \eqref{4_C03} limits the number of users associated with a access point to a maximum of $F_n$. Constraint \eqref{4_C04} specifies that the cache decision variables are binary. Constraint \eqref{4_C05} indicates that the user association decision variables are binary as well. Constraint \eqref{4_C06} specifies the transmit power constraint for FAP $n$, where the maximum allowable transmit power is denoted by $p_{n}^{max}$. Finally, constraint \eqref{4_C07} asserts that the power allocation variables are positive. Constraint \eqref{4_C08} defines the power constraint of the CP, with a maximum transmit power limit of $p_{0}^{max}$. Finally, constraint \eqref{4_C09} asserts that the power allocation variables for the CP are positive.

The formulated problem is a non-convex MINLP problem. The user association variable $x_{n,k}$ and the cache placement variable $c_{n,f}$ are binary, which introduces integer constraints in \eqref{4_C01}, \eqref{4_C02}, and \eqref{4_C03}. Moreover, the objective function presented in \eqref{4_ProblemFormulation1} is non-convex, making the optimization problem inherently combinatorial. While techniques such as exhaustive search can guarantee finding the global optimum, they are typically associated with exponentially increasing computational complexity.
\section{Proposed low complexity Soulution}\label{AO}
To overcome aforementioned issue, the problem \eqref{4_ProblemFormulation1} is decomposed into two subproblems, namely 1) the user association and cache placement subproblem, and 2) the power allocation subproblem. These subproblems are then solved alternately in an iterative manner. It should be noted that the constant signaling overhead $D_{\text{ov}}$, introduced in the delay model, is omitted in the following objective function formulations. Since $D_{\text{ov}}$ is a constant term independent of the optimization variables, its exclusion simplifies the mathematical derivation without affecting the determination of the optimal solution. In Section \ref{mccormick} and Section \ref{IV}, a low-complexity joint user association and cache placement algorithm based on the McCormick envelope theory and Lagrangian partial relaxation method is proposed, along with a low-complexity power allocation algorithm based on SCA. 
\subsection{Proposed algorithm for user association and caching placement}\label{mccormick}
In this subsection, we propose the algorithms that solve the user association and caching placement subproblem. Given that the user association and cache placement subproblems are NP-hard, the complexity is exceedingly high. To reduce the problem’s complexity, this subsection proposes a distributed algorithm. First, leveraging McCormick envelope theory, the optimization problem is equivalently transformed. Next, the transformed problem is solved using a Lagrangian partial relaxation method, breaking it down into several subproblems. It is evident that the cache variables $c_{n,f}$ and user association variables $x_{n,k}$ are coupled in the objective function \eqref{4_C0}, making it challenging to solve. To decouple the user association and cache variables in the problem \eqref{4_ProblemFormulation1}, new variable $z_{f,n}^k$ is introduced, where $z_{f,n}^k = (1 - c_{n,f})x_{n,k}$. Consequently, the optimization subproblem for a fixed power allocation can be reformulated as follows:
\begin{subequations}\label{4_ProblemFormulation2}
	\begin{align}
		\displaystyle{\mathop {\min }\limits_{\{ \bf{x},\bf{c},\bf{z} \} }} &~~ \displaystyle{ \frac{1}{K}(\sum\limits_{n = 1}^N \sum\limits_{k = 1}^K  (x_{n,k} D_{n,k}^F+z_{f,n}^k D_{n,k}^B) }+ \sum\limits_{k = 1}^K x_{0,k} D_{0,k} )\label{4_C00}\\ 
		\text{s.t.}&~~ \eqref{C01order}-\eqref{4_C05} ,\label{4_C001}\\
		&~~z_{f,n}^k=(1-c_{n,f})x_{n,k}, \forall n \in\mathcal{N}^+,\forall k \in\mathcal{K},\forall f\in\mathcal{F}.\label{4_C002}
	\end{align}
\end{subequations}
Here, $\mathcal{N}^+ = \mathcal{N} \setminus \{0\}$ denotes the set of FAPs excluding the CP, \eqref{4_C002} represents a non-convex constraint. By utilizing the McCormick envelope theory \cite{liberti2006exact} to relax the constraint, it is given by the following equations:
\begin{eqnarray}\label{zm1}
	z_{f,n}^k \geq x_{n,k}-c_{n,f},\forall n \in\mathcal{N}^+,\forall k \in\mathcal{K},\forall f\in\mathcal{F},
\end{eqnarray}
\begin{eqnarray}\label{zm2}
	z_{f,n}^k \geq 0,\forall k \in\mathcal{K},\forall f\in\mathcal{F},
\end{eqnarray}
\begin{eqnarray}\label{zm3}
	z_{f,n}^k \leq x_{n,k},\forall k \in\mathcal{K},\forall f\in\mathcal{F},
\end{eqnarray}
\begin{eqnarray}\label{zm4}
	z_{f,n}^k \leq 1-c_{n,f},\forall k \in\mathcal{K},\forall f\in\mathcal{F}.
\end{eqnarray}
Given the binary and discrete nature of the optimization variables $c_{n,f}$ and $x_{n,k}$, it can be rigorously proven that the equality $z_{f,n}^k = (1 - c_{n,f})x_{n,k}$ is equivalent to the constraint conditions \eqref{zm1} and \eqref{zm4}, as shown in Table \ref{4_T:equation}.
\begin{table}[!htb]
	\vspace{-0.2cm}\small
	\renewcommand{\arraystretch}{1.3}
	\caption{Equivalent Transformation between Equality \eqref{4_C002} and Convex Relaxation Conditions}
	\label{4_T:equation}
	\begin{center}\vspace{-0.4cm}
		\begin{tabular}{|p{0.4cm}<{\centering}|p{0.4cm}<{\centering}|p{0.4cm}<{\centering}|p{2.6cm}<{\centering}|p{2.5cm}<{\centering}|}
			\hline
			$x_{n,k}$&$c_{n,f}$&$z_{f,n}^k$&$\max(x_{n,k}-c_{n,f},0)$&$\min(x_{n,k},1-c_{n,f})$\\ \hline
			0&0&0&0&0  \\\hline
			0&1&0&0&0 \\\hline
			1&0&1&1&1 \\\hline
			1&1&0&0&0 \\\hline
		\end{tabular}
	\end{center}
	\vspace{-0.2cm}
\end{table}

Therefore, the optimization problem \eqref{4_ProblemFormulation2} can be further expressed as:
\begin{subequations}\label{4_ProblemFormulation3}
	\begin{align}
		\displaystyle{\mathop {\min }\limits_{\{ \bf{x},\bf{c},\bf{z}  \} }}&~~\displaystyle{ \frac{1}{K}(\sum\limits_{n = 1}^N \sum\limits_{k = 1}^K  (x_{n,k} D_{n,k}^F+z_{f,n}^k D_{n,k}^B) + \sum\limits_{k = 1}^K x_{0,k} D_{0,k} )}\label{4_C000}\\ 
		\text{s.t.}&~~\eqref{C01order}-\eqref{4_C05},\eqref{zm1}-\eqref{zm4}.\label{4_C0001}
	\end{align}
\end{subequations}

To solve the new optimization problem \eqref{4_ProblemFormulation3}, the Lagrangian partial relaxation method is employed \cite{boyd2004convex}. Specifically, the constraints \eqref{zm1}, \eqref{zm3}, and \eqref{zm4} are relaxed, and a set of dual Lagrangian multipliers is introduced:
\begin{eqnarray}\label{mu1}
	\lambda_{f,n}^k \geq 0, \forall n \in\mathcal{N}^+,\forall k \in\mathcal{K},\forall f\in\mathcal{F},
\end{eqnarray}
\begin{eqnarray}\label{lam1}
	\mu_{f,n}^k \geq 0, \forall n \in\mathcal{N}^+,\forall k \in\mathcal{K},\forall f\in\mathcal{F},
\end{eqnarray}
\begin{eqnarray}\label{ps1}
	\psi_{f,n}^k \geq 0, \forall n \in\mathcal{N}^+,\forall k \in\mathcal{K},\forall f\in\mathcal{F}.
\end{eqnarray}
Thus, the Lagrangian function can be expressed as:
\begin{equation}\label{Lagrange1}
	\begin{split}
&L({\bm{\lambda},\bm{\mu},\bm{\psi},\bf{x},\bf{c},\bf{z}})=\\
&	\frac{1}{K}(\sum\limits_{n = 1}^N \sum\limits_{k = 1}^K (x_{n,k} D_{n,k}^F+z_{f,n}^k D_{n,k}^B) + \sum\limits_{k = 1}^K x_{0,k} D_{0,k})\\
&+\sum\limits_{n = 1}^N \sum\limits_{k = 1}^K \sum\limits_{f = 1}^F [\mu_{f,n}^k(x_{n,k}-c_{n,f}-z_{f,n}^k)+\lambda_{f,n}^k (z_{f,n}^k-x_{n,k})\\
& +\psi_{f,n}^k(z_{f,n}^k+c_{n,f}-1)].
	\end{split}
\end{equation}

The dual problem corresponding to the original optimization formulation can be derived by introducing dual variables, leading to the following dual optimization problem:
\begin{subequations}\label{dual}
	\begin{align}
		\displaystyle{\mathop{\max }\limits_{\{ \bm{\lambda},\bm{\mu},\bm{\psi} \} } \mathop {\min }\limits_{\{ \bf{x},\bf{c},\bf{z} \} }} &~~ \displaystyle{L({\bm{\lambda},\bm{\mu},\bm{\psi},\bf{x},\bf{c},\bf{z}}) }\label{4_C0000}\\ 
		\text{s.t.}&~~\eqref{C01order}-\eqref{4_C05},\eqref{zm2},\eqref{mu1}-\eqref{ps1}.\label{4_C00001}
	\end{align}
\end{subequations}
In this formulation, the objective function of the dual problem and is given by
\begin{eqnarray}\label{pxz}
	L({\bm{\lambda},\bm{\mu},\bm{\psi},\bf{x},\bf{c},\bf{z}})=f(\bf{x})+g(\bf{c})+h(\bf{z}),
\end{eqnarray}
Here, the objective functions $f(\bf{x})$, $g(\bf{c})$, and $h(\bf{z})$ correspond to the subproblems P1, P2, and P3, respectively. Moreover, the feasible region of the problem \eqref{dual} is decomposable into three distinct regions, namely \{\eqref{C01order}, \eqref{4_C02}, \eqref{4_C03}, \eqref{4_C05}\}, \{\eqref{4_C01}, \eqref{4_C04}\} and \{\eqref{zm2}\}. As a result, the problem \eqref{dual} is decomposed into three separate subproblems, denoted as P1, P2, and P3. These subproblems can be formulated as follows:
\begin{subequations}\label{P1}
	\begin{align}
		\begin{split}
		P1: \mathop {\min }\limits_{\{ \bf{x} \} } &~~\frac{1}{K}(\sum\limits_{n = 1}^N \sum\limits_{k = 1}^K x_{n,k} D_{n,k}^F + \sum\limits_{k = 1}^K x_{0,k} D_{0,k})\\
		&+\sum\limits_{n = 1}^N \sum\limits_{k = 1}^K \sum\limits_{f = 1}^F (\mu_{f,n}^k x_{n,k}-\lambda_{f,n}^k x_{n,k})\label{4_C00000}\end{split}\\ 
		\text{s.t.}&~~\eqref{C01order}, \eqref{4_C02}, \eqref{4_C03},\eqref{4_C05}.\label{4_C000001}
	\end{align}
\end{subequations}
\begin{subequations}\label{P2}
	\begin{align}
		\displaystyle{P2: \mathop {\max }\limits_{\{ \bf{c} \} }} &~~ \displaystyle{ \sum\limits_{n = 1}^N \sum\limits_{k = 1}^K \sum\limits_{f = 1}^F (\mu_{f,n}^k c_{n,f}-\psi_{f,n}^k c_{n,f})}\label{4_C000000}\\ 
		\text{s.t.} &~~\eqref{4_C01}, \eqref{4_C04}.\label{4_C0000001}			 
	\end{align}
\end{subequations}
\begin{subequations}\label{P3}
	\begin{align}
		\begin{split}
		\displaystyle{P3: \mathop {\min }\limits_{\{ \bf{z} \} }} &~~\frac{1}{K}\sum\limits_{n = 1}^N \sum\limits_{k = 1}^K z_{f,n}^k D_{n,k}^B\\ 
		&+\sum\limits_{n = 1}^N \sum\limits_{k = 1}^K \sum\limits_{f = 1}^F (\lambda_{f,n}^k z_{f,n}^k+\psi_{f,n}^k z_{f,n}^k-\mu_{f,n}^k z_{f,n}^k)
		\label{4_C0000000}\end{split}\\ 
		\text{s.t.} &~~\eqref{zm2}.\label{4_C00000001}				 
	\end{align}
\end{subequations}

After decomposition, the joint optimization problem is effectively reduced to separate optimization tasks, removing the coupling between user association and cache placement variables. It can be demonstrated that these three subproblems are integer programming problems given a specific set of Lagrange multipliers. Specifically, P1, which involves user association, can be addressed using the Hungarian algorithm \cite{kuhn1955hungarian}. Meanwhile, P2 which involve cache placement and related decisions, can be efficiently solved via integer linear programming \cite{korte2011combinatorial}. For P3, \( z_{f,n}^k \) is set to 1 if \( \frac{1}{K} D_{n,k}^B + \lambda_{f,n}^k + \psi_{f,n}^k - \mu_{f,n}^k < 0 \), and 0 otherwise. Upon solving these subproblems, the optimal values for $\bf{x}$, $\bf{c}$, and $\bf{z}$ are obtained. Subsequently, the dual problem is addressed using the subgradient method, where the subgradients of the dual function are computed as follows:
\begin{equation}\label{subgradientmu1}
	\begin{split}
	\nabla \mu_{f,n}^k(t)=&x_{n,k}(t)-c_{n,f}(t)-z_{f,n}^k(t), \\
	&\forall n \in\mathcal{N}^+,\forall k \in\mathcal{K},\forall f\in\mathcal{F},
    \end{split}
\end{equation}
\begin{equation}\label{subgradientlam1}
	\begin{split}
	\nabla \lambda_{f,n}^k(t)=&z_{f,n}^k(t)-x_{n,k}(t),\\ 
	&\forall n \in\mathcal{N}^+,\forall k \in\mathcal{K},\forall f\in\mathcal{F},
\end{split}
\end{equation}
\begin{equation}\label{subgradientps1}
	\begin{split}
	\nabla \psi_{f,n}^k(t)=&z_{f,n}^k(t)+c_{n,f}(t)-1,\\
	& \forall n \in\mathcal{N}^+,\forall k \in\mathcal{K},\forall f\in\mathcal{F},
\end{split}
\end{equation}
In the $(t+1)$th iteration, the subgradients $\nabla \mu_{f,n}^k(t)$, $\nabla \lambda_{f,n}^k(t)$ and $\nabla \psi_{f,n}^k(t)$ are utilized to update three dual variables. Consequently, three update expressions for the dual variables at iteration $(t+1)$ can be formulated as follows:
\begin{equation}\label{updatemu1}
\begin{split}
\mu_{f,n}^k(t+1)=&{\left\lbrack \mu_{f,n}^k(t)+\xi(t)\times \nabla \mu_{f,n}^k(t)\right\rbrack}^{+},\\
&\forall n \in\mathcal{N}^+,\forall k \in\mathcal{K},\forall f\in\mathcal{F},
\end{split}
\end{equation}
\begin{equation}\label{updatelam1}
\begin{split}
\lambda_{f,n}^k(t+1)=&{\left\lbrack \lambda_{f,n}^k(t)+\xi(t)\times \nabla \lambda_{f,n}^k(t)\right\rbrack}^{+}\\
&\forall n \in\mathcal{N}^+,\forall k \in\mathcal{K},\forall f\in\mathcal{F},
\end{split}
\end{equation}
\begin{equation}\label{updateps1}
	\begin{split}
	\psi_{f,n}^k(t+1)=&{\left\lbrack \psi_{f,n}^k(t)+\xi(t)\times \nabla \psi_{f,n}^k(t)\right\rbrack}^{+},\\
	&\forall n \in\mathcal{N}^+,\forall k \in\mathcal{K},\forall f\in\mathcal{F},
\end{split}
\end{equation}
where ${\left\lbrack x \right\rbrack}^{+}=\max\{x,0\}$ denotes the non-negative part of $x$, and \( \xi(t) = \frac{0.01}{\sqrt{t}} \) represents the positive step size at the $(t+1)$th iteration. It is important to note that this step size is non-accumulative and diminishes over time. As demonstrated in \cite{bertsekas1997nonlinear}, such a step size ensures that the algorithm converges to the optimal solution.

\begin{algorithm}[!htb]
	\small
    \caption{Joint User Association and Cache Placement Algorithm}
    \label{4_JACPM}
    \begin{algorithmic}[1]
        \STATE Set \( \mu_{f,n}^k(1) = 0.1 \) for files \( f \) requested by user \( k \), else 0; \( \lambda_{f,n}^k(1) = 0 \), \( \psi_{f,n}^k(1) = 0 \), $W(1)=0$, \( t = 1 \), \( t_{\text{max}} = 20 \), \( \varepsilon = 10^{-3} \). Compute initial decoding order and \( F_{\text{push}} \).
        \STATE \textbf{While} \( t \leq t_{\text{max}} \) and \( |W(t) - W(t-1)| \geq \varepsilon \) \textbf{do}
        \STATE~~Solve subproblem P1 for \( \mathbf{x}(t) \).
	    \STATE~~Solve subproblem P2 for \( \mathbf{c}(t) \).
	    \STATE~~Solve subproblem P3 for \( \mathbf{z}(t) \).
		\STATE~~Update decoding order and \( F_{\text{push}} \) based on \( \mathbf{x}(t) \) and \( \mathbf{c}(t) \).
        \STATE~~Compute the objective function value \( W(t) =\)\\~~\( \frac{1}{K} ( \sum\limits_{n=1}^N \sum\limits_{k=1}^K (x_{n,k} D_{n,k}^F + z_{f,n}^k D_{n,k}^B) + \sum\limits_{k=1}^K x_{0,k} D_{0,k} ) \).
        \STATE~~Update multipliers via \eqref{updatemu1}-\eqref{updateps1} with \( \xi(t) = \frac{0.01}{\sqrt{t}} \).
        \STATE~~\( t = t + 1 \)
        \STATE \textbf{End While}
        \STATE \textbf{Output}: \( \mathbf{x}^* = \mathbf{x}(t) \), \( \mathbf{c}^* = \mathbf{c}(t) \)
    \end{algorithmic}
\end{algorithm}

The proposed algorithm, outlined in Algorithm \ref{4_JACPM}, addresses the joint user association and caching placement optimization problem. Its overall computational complexity is evaluated under the assumption that subproblems P1, P2, and P3 are executed in parallel within each iteration. The analysis considers the per-iteration complexity and the number of iterations required for convergence. P1 is solved using an Hungarian algorithm, optimizing user assignments across \(N+1\) nodes. Its complexity is \(\mathcal{O}(K^3)\), driven by the cubic scaling of the Hungarian method with the number of UEs \(K\), which dominates other computations like decoding order and interference calculations. P2 optimizes caching decisions for each of the \(N\) FAPs via integer linear programming. Using practical solvers (e.g., interior-point methods for relaxed problems), the complexity per FAP is \(\mathcal{O}(F^{3.5})\), where \(F\) is the number of files. Across \(N\) FAPs, the total complexity is \(\mathcal{O}(N F^{3.5})\). P3 optimizes auxiliary variables for FAP-UE-file triplets, with a complexity of \(\mathcal{O}(N K F)\), as it scales linearly with the number of FAPs \(N\), UEs \(K\), and files \(F\). When P1, P2, and P3 are solved in parallel, the per-iteration complexity is governed by the most computationally intensive subproblem. The algorithm converges in \(\mathcal{O}(1 / \varepsilon)\) iterations. Hence, the total time complexity is: \(\mathcal{O} \left( \frac{1}{\varepsilon} \cdot \max \left( K^3, N F^{3.5}, N K F \right) \right)\). An exhaustive search would yield a complexity of \(\mathcal{O}((N+1)^K)\), exponentially scaling with \(K\). In contrast, the proposed algorithm’s polynomial complexity offers substantial efficiency, making it practical for large-scale wireless networks.

\subsection{Proposed power allocation algorithm}\label{IV}
In this subsection, we address the power allocation subproblem under fixed user association and cache placement, as determined by the solutions to the subproblems discussed in Section~\ref{mccormick}. It is important to highlight that the model considered here does not account for cross-layer interference between the CP and FAPs, focusing only on intra-cell and inter-cell interference. For each layer, the power allocation coefficients within that layer do not influence the performance of other layers. The objective is to minimize the average delay in content delivery across a wireless network comprising a CP and \( N \) FAPs, while adhering to power constraints and accounting for interference and caching effects. The power allocation coefficients can be iteratively optimized during each iteration of the proposed joint user association and cache placement algorithm, enabling the simultaneous improvement of both aspects. The original problem formulated in \eqref{4_ProblemFormulation1} under fixed user association and cache placement is given by:
\begin{subequations}\label{4_power}
	\begin{align}
		\displaystyle{\mathop {\min }\limits_{  \{p_{n,k} \} }} &~~\displaystyle{ \frac{1}{K} (\sum\limits_{n = 1}^N \sum\limits_{k\in\mathcal{K}_n} {D_{n,k}}+\sum\limits_{k \in\mathcal{K}_0} D_{0,k})}\label{4_P00}\\ 
		\text{s.t.}&~~\sum\limits_{k \in \mathcal{K}_n}  {p_{n,k}} \leq p_{n}^{max},\forall n \in\mathcal{N}^+,\label{4_P001}\\
		    &~~{p_{n,k}} \geq 0, \forall n \in\mathcal{N}^+,\forall k \in\mathcal{K}_n,\label{4_P002}\\
		&~~p_{0,k}+\sum_{{{f \in \mathcal{F}^{\text{push}}}}}p_{0,f}\leq p_{0}^{max}, \label{4_P003}\\
		&~~{p_{0,k}} \geq 0,{p_{0,f}} \geq 0, \forall k \in\mathcal{K}_0, \forall f \in \mathcal{F}^{\text{push}}. \label{4_P004}
	\end{align}
\end{subequations}
This problem is non-convex due to the interference-coupled rate expressions. To make the problem tractable, we decompose it into two interdependent subproblems: FAP power allocation with interference coordination, and CP power allocation with a push-delivery tradeoff.

The FAPs power allocation subproblem must consider inter-FAP interference while satisfying the per-FAP power constraints. This subproblem is formulated as follows:
\begin{subequations}\label{4_powerfap}
	\begin{align}
		\displaystyle{\mathop {\min }\limits_{\{  p_{n,k} \} }} &~~ \displaystyle{ \sum\limits_{n = 1}^N\sum\limits_{k\in\mathcal{K}_n}\sum\limits_{f = 1}^F \frac{r_{k,f}  s_f }{R_{n,k}}}\label{4_P0}\\ 
		\text{s.t.} &~~ \sum\limits_{k \in \mathcal{K}_n}  {p_{n,k}} \leq p_{n}^{max},\forall n \in\mathcal{N}^+,\label{4_P01}\\
		&~~{p_{n,k}} \geq 0,\forall n \in\mathcal{N}^+, \forall k \in\mathcal{K}_n.\label{4_P02}
	\end{align}
\end{subequations}

The CP power allocation subproblem balances direct content delivery to users and proactive content pushing to FAPs. This subproblem is expressed as follows:
\begin{subequations}\label{4_powerccp}
	\begin{align}
		\mathop{\min}\limits_{\{p_{0,k},p_{0,f}\}} & ~~  \sum\limits_{k \in\mathcal{K}_0} \sum\limits_{f = 1}^F \frac{r_{k,f}  s_f }{R_{0,k}}+{ \sum\limits_{n = 1}^N \sum\limits_{k\in\mathcal{K}_n^f}\sum\limits_{f \in \mathcal{F}^{\text{push}}} \frac{ r_{k,f} s_f }{R_{0,n}^f}}, \label{CP_obj}\\ 
		\text{s.t.} &~~ p_{0,k} + \sum_{f \in \mathcal{F}^{\text{push}}} p_{0,f} \leq p_0^{max},\label{CP_C1}\\
		&~~p_{0,k} \geq 0, p_{0,f} \geq 0,  \forall k \in \mathcal{K}_0, \forall f \in \mathcal{F}^{\text{push}}.\label{CP_C3}
	\end{align}
\end{subequations}
Solving the two subproblems provides the solution to the original power allocation problem \eqref{4_power}. The power allocation problem for the CP is conceptually similar to that for the FAPs.

Problem \eqref{4_powerfap} is non-convex due to its non-convex objective function in \eqref{4_P0}, although the constraints are convex. This type of problem is NP-hard, and a global optimal solution can theoretically be found using the Branch-and-Bound (BB) algorithm \cite{wu2022non}. However, the exhaustive search nature of this method results in high computational complexity and inefficiency. To overcome these challenges, we propose a low-complexity solution based on the SCA technique, which effectively addresses problem \eqref{4_powerfap}.

First, we propose an equivalent problem for \eqref{4_powerfap}, followed by approximating the non-convex constraints using a first-order approximation to transform them into convex constraints. Next, an iterative algorithm is proposed by applying the SCA algorithm to the transformed problem. Specifically, in the $l$-th iteration of the proposed algorithm, problem \eqref{4_powerfap} is equivalently transformed into the following problem:
\begin{subequations}\label{4_powerequ}
	\begin{align}
		\displaystyle{\mathop {\min }\limits_{\{  p_{n,k},\hat{D},\tau_{n,k},\atop \upsilon_{n,k}, \varsigma_{n,k} \} }} &~~ \displaystyle{ {{\hat{D}}}} \label{4_e0} \\ 
		\text{s.t.} &~~ \sum\limits_{n = 1}^N\sum\limits_{k\in\mathcal{K}_n} \sum\limits_{f = 1}^F (\frac{ r_{k,f} s_f }{\tau_{n,k}})\leq {{\hat{D}}}, \label{4_e01} \\
		  &~~ \tau_{n,k} \leq B \log_2 (1+\upsilon_{n,k}), \label{4_e02} \\
	     &~~ \upsilon_{n,k} \leq   \frac{ {{\left| {{h_{n,k}}} \right|}^2}{p_{n,k}} }{ \varsigma_{n,k}}, \label{4_e03}	\\	
		\begin{split}
		&~~  {{\left| {{h_{n,k}}} \right|}^2} \sum_{i=k+1}^{t}{p_{n,i}}+\sum_{j = 1,j \ne n}^N {{\left| {{h_{j,k}}} \right|}^2} p_j \\
		&~~ +\sigma^2 \leq   { \varsigma_{n,k}}, \label{4_e04} \end{split}\\
		&~~\eqref{4_P01},\eqref{4_P02},\label{4_e05}
	\end{align}
\end{subequations}
where $n \in\mathcal{N}^+, f\in\mathcal{F}, k \in\mathcal{K}$. $(p_{n,k}^\star,\tau_{n,k}^\star,\upsilon_{n,k}^\star, \varsigma_{n,k}^\star)$ and ${\hat{D}}^\star$ denote the optimal solution and corresponding optimal value, respectively.

\textbf{Lemma 1} (Equivalence between problem \eqref{4_powerfap} and problem \eqref{4_powerequ}): Problem \eqref{4_powerfap} is equivalent to problem \eqref{4_powerequ}.

\emph{Proof}: It is clear that problem \eqref{4_powerfap} qualifies as a non-convex optimization problem, primarily due to its non-convex nature of the objective function. To aid in solving the problem, we introduce a new variable $\hat{D}$, allowing the original problem to be equivalently reformulated as follows:
\begin{subequations}\label{4_powerfapDDD}
	\begin{align}
		\displaystyle{\mathop {\min }\limits_{\{  p_{n,k}, {\hat{D} \} }}} &~~ \displaystyle{    {\hat{D}}   }\label{4_PD0}\\ 
		\text{s.t.}&~~  \sum\limits_{n = 1}^N\sum\limits_{k\in\mathcal{K}_n} \sum\limits_{f = 1}^F (\frac{ r_{k,f} s_f }{R_{n,k}} )\leq {{\hat{D}}},\label{4_PD01}\\
		&~~\eqref{4_P01},\eqref{4_P02}.\label{4_PD02}
	\end{align}
\end{subequations}
The equivalence between problem \eqref{4_powerfapDDD} and problem \eqref{4_powerfap} is straightforwardly proved as all constraints in \eqref{4_powerfapDDD} remain equal under optimal conditions.

By introducing new variables $\bm{\tau} \triangleq (\tau_{n,k})_{n \in\mathcal{N},k \in\mathcal{K}}$ and focusing on constraint \eqref{4_PD01}, it is equivalent to the following two constraints:
\begin{equation}\label{hatd}
	 \sum\limits_{n = 1}^N\sum\limits_{k\in\mathcal{K}_n} \sum\limits_{f = 1}^F (\frac{ r_{k,f} s_f }{\tau_{n,k}} )\leq {{\hat{D}}},
\end{equation}
\begin{equation}\label{tau}
	\tau_{n,k} \leq R_{n,k},
\end{equation}
where $n \in\mathcal{N}^+, f\in\mathcal{F}, k \in\mathcal{K}$. Next, by introducing two new variables $\bm{\upsilon} \triangleq (\upsilon_{n,k})_{n \in\mathcal{N},k \in\mathcal{K}}$ and $\bm{\varsigma} \triangleq (\varsigma_{n,k})_{n \in\mathcal{N},k \in\mathcal{K}}$, equation \eqref{tau} can be equivalently rewritten as:
\begin{equation}\label{tauupsilon}
	\tau_{n,k} \leq B \log_2 (1+\upsilon_{n,k}),
\end{equation}
\begin{equation}\label{upsilon}
	\upsilon_{n,k} \leq   \frac{ {{\left| {{h_{n,k}}} \right|}^2}{p_{n,k}} }{ \varsigma_{n,k}},
\end{equation}
\begin{equation}\label{varsigma}
	{ {{\left| {{h_{n,k}}} \right|}^2} \sum_{i=k+1}^{t}{p_{n,i}}+\sum_{j = 1,j \ne n}^N {{\left| {{h_{j,k}}} \right|}^2} p_j+\sigma^2} \leq   { \varsigma_{n,k}} .
\end{equation} 
With the above analysis, the equivalence transformation of problem \eqref{4_powerfap} is easily obtained when the constraints \eqref{hatd}, \eqref{tauupsilon}--\eqref{varsigma} hold as equalities \cite{tran2012fast}. The proof is completed.\hfill$\blacksquare$ 

All constraints in problem \eqref{4_powerequ} are convex except for \eqref{4_e03}. First, consider the transformation of constraint \eqref{4_e03}, where the term $\upsilon_{n,k}\varsigma_{n,k}$ can be rewritten as follows:
\begin{equation}\label{upsilonvarsigma}
	\upsilon_{n,k}\varsigma_{n,k}=\frac{1}{4}[(\upsilon_{n,k}+\varsigma_{n,k})^2-(\upsilon_{n,k}-\varsigma_{n,k})^2].
\end{equation}

After transforming the bilinear term, the resulting expression takes the form of the difference of two convex functions within the brackets. This non-convex formulation can be efficiently addressed using SCA algorithms \cite{hanif2015minorization}. The core idea behind SCA is to approximate the objective function with a first-order Taylor series, which provides a convex lower or upper bound. In equation \eqref{upsilonvarsigma}, the second term in the brackets is replaced by its convex lower bound. By applying the SCA method, the Taylor series expansion at the point $(\upsilon_{n,k}^{(\Phi)}, \varsigma_{n,k}^{(\Phi)})$ is derived as follows:
\begin{equation}\label{upsilonvarsigmaphi}
	\begin{split}
	f(\upsilon_{n,k},\varsigma_{n,k})&=(\upsilon_{n,k}-\varsigma_{n,k})^2\\
	&\geq (\upsilon_{n,k}^{(\Phi)}-\varsigma_{n,k}^{(\Phi)})^2\\
	&+2(  \upsilon_{n,k}^{(\Phi)}-\varsigma_{n,k}^{(\Phi)}      )[  (  \upsilon_{n,k}-\upsilon_{n,k}^{(\Phi)}  )  -( \varsigma_{n,k}- \varsigma_{n,k}^{(\Phi)}     )    ] \\
	&\triangleq g( \upsilon_{n,k}, \varsigma_{n,k},  \upsilon_{n,k}^{(\Phi)}, \varsigma_{n,k}^{(\Phi)}   ).
	\end{split}
\end{equation}
At the point $(\upsilon_{n,k}, \varsigma_{n,k}) = (\upsilon_{n,k}^{(\Phi)}, \varsigma_{n,k}^{(\Phi)})$, the affine approximation aligns with the function $f(\upsilon_{n,k}, \varsigma_{n,k})$. During the iterative process, the variable values $(\upsilon_{n,k}, \varsigma_{n,k})$ are updated to $(\upsilon_{n,k}^{(\Phi+1)}, \varsigma_{n,k}^{(\Phi+1)})$, satisfying the following conditions:
\begin{equation}\label{fffupsilonvarsigma}
	\begin{split}
&f(\upsilon_{n,k},\varsigma_{n,k})\geq g( \upsilon_{n,k}, \varsigma_{n,k},  \upsilon_{n,k}^{(\Phi)}, \varsigma_{n,k}^{(\Phi)}   ),\\
&f(\upsilon_{n,k}^{(\Phi)},\varsigma_{n,k}^{(\Phi)})= g( \upsilon_{n,k}^{(\Phi)}, \varsigma_{n,k}^{(\Phi)},  \upsilon_{n,k}^{(\Phi)}, \varsigma_{n,k}^{(\Phi)}   ),\\
&\nabla f(\upsilon_{n,k},\varsigma_{n,k}) \mid_{(\upsilon_{n,k}^{(\Phi)},\varsigma_{n,k}^{(\Phi)})}=\\
& \nabla g( \upsilon_{n,k}, \varsigma_{n,k},  \upsilon_{n,k}^{(\Phi)},\varsigma_{n,k}^{(\Phi)}) \mid_{(\upsilon_{n,k}^{(\Phi)},\varsigma_{n,k}^{(\Phi)})},
\end{split}
\end{equation}
where $\nabla f$ denotes the function gradient. Consequently, the problem \eqref{4_powerequ} can be transformed into
\begin{subequations}\label{powerequtran}
	\begin{align}
		\displaystyle{\mathop {\min }\limits_{\{  \bf{p},\hat{D},\bm{\tau},\bm{\upsilon}, \bm{\varsigma} \} }} &~~\displaystyle{ {{\hat{D}}}} \label{et0} \\ 
		\begin{split}
		\text{s.t.} & ~~(\upsilon_{n,k}+\varsigma_{n,k})^2-g( \upsilon_{n,k}, \varsigma_{n,k},  \upsilon_{n,k}^{(\Phi)}, \varsigma_{n,k}^{(\Phi)}) \\
		&~~\leq   4 {{\left| {{h_{n,k}}} \right|}^2}{p_{n,k}},\label{et1}\end{split}\\
		& ~~\eqref{4_P01},\eqref{4_P02},\eqref{4_e01},\eqref{4_e02},\eqref{4_e04}.\label{et2}
		\end{align}
\end{subequations}

The optimization problem \eqref{powerequtran} is convex and can be efficiently solved for the $\Phi$-th iteration, given the initial values $\upsilon_{n,k}^{(\Phi)}$ and $\varsigma_{n,k}^{(\Phi)}$ \cite{boyd2004convex}. As established in \cite{wang2019user}, the SCA-based approach to solving problem \eqref{4_powerfap} converges to a local optimum. With an increasing number of iterations, the algorithm approaches KKT points of the original problem \eqref{4_powerequ}.

Similar to Lemma 1, the CP power allocation subproblem \eqref{4_powerccp} can be equivalently transformed into the optimization problem \eqref{4_powereCP} through variable substitution and constraint reformulation. The structural similarity between problems \eqref{4_powerfap} and \eqref{4_powerccp} allows the SCA methodology developed for FAPs to be directly applied to the CP case. Specifically, the non-convex terms in \eqref{CP_obj} can be handled using the same convex approximation technique presented in \eqref{upsilonvarsigma}-\eqref{fffupsilonvarsigma}. Therefore, we omit the detailed derivation for brevity. The CP optimization problem can be transformed into

\begin{subequations}\label{4_powereCP}
	\begin{align}
		\displaystyle{\mathop{\min }\limits_{\{ \substack{ p_{0,k}, p_{0,f}, \hat{D}_{CP},  \tau_{0,k},\\ \tau_{0,n}^f, \upsilon_{0,k},  \upsilon_{0,n}^f, \varsigma_{0,k}, \varsigma_{0,n}^f } \}}} &~~ \displaystyle{ {{\hat{D}}_{CP}}} \label{CP4_e0} \\ 
		\begin{split}
		\text{s.t.} & \sum_{k \in \mathcal{K}_0} \sum_{f=1}^F \frac{r_{k,f} s_f}{\tau_{0,k}} + \\
		&\sum\limits_{n = 1}^N\!\sum\limits_{k\in\mathcal{K}_n^f}\!\sum_{f \in \mathcal{F}^{\text{push}}} \frac{r_{k,f} s_f}{\tau_{0,n}^f} \leq {{\hat{D}}_{CP}}, \label{CP4_e01} \end{split}\\
		& \tau_{0,k} \leq B \log_2(1 + \upsilon_{0,k}), \label{CP4_e02} \\
	    & \tau_{0,n}^f \leq B \log_2(1 + \upsilon_{0,n}^f), \label{CP4_e03} \\
		& \upsilon_{0,k} \leq \frac{|h_{0,k}|^2 p_{0,k}}{\varsigma_{0,k}}, \label{CP4_e04}	\\
		& \varsigma_{0,k} \geq |h_{0,k}|^2 \sum_{f \in \mathcal{F}^{\text{push}}} p_{0,f} + \sigma^2, \label{CP4_e05}	\\
		& \upsilon_{0,n}^f \leq \frac{L_{0,n}^{-1} p_{0,f}}{\varsigma_{0,n}^f}, \label{CP4_e06}	\\	
		& \varsigma_{0,n}^f \geq L_{0,n}^{-1} \sum\limits_{i=f+1}^{|\mathcal{F}^{\text{push}}|} p_{0,i} + \sigma^2, \label{CP4_e07}	\\
		&\eqref{CP_C1},\eqref{CP_C3},\label{CP4_e08}
	\end{align}
\end{subequations}
where $n \in\{{0}\}, f\in\mathcal{F}, k \in\mathcal{K}$. $(p_{0,k}^\star, p_{0,f}^\star)$, $(\tau_{0,k}^\star, \tau_{0,n}^{f\star})$, $(\upsilon_{0,k}^\star, \upsilon_{0,n}^{f\star})$, $(\varsigma_{0,k}^\star, \varsigma_{0,n}^{f\star})$ and $\hat{D}_{CP}^\star$ denote the optimal solution and corresponding optimal value, respectively.
\begin{algorithm}[htb]
	\small
    \caption{SCA-based Power Allocation Algorithm}
    \label{4_IRLS}
    \begin{algorithmic}[1]
		\STATE Initialize $\Phi = 0$, FAP powers $\mathbf{P}_{\text{FAP}}^{(0)}$, CP powers $\mathbf{P}_{\text{CP}}^{(0)}$, and auxiliary variables $(\upsilon_{n,k}^{(0)}, \varsigma_{n,k}^{(0)})$.
        \STATE \textbf{FAP Power Allocation:}
        \STATE \quad \textbf{While} $\Phi < \Phi_{\max}$ and not converged \textbf{do}
        \STATE \quad \quad Update $(\upsilon_{n,k}^{(\Phi)}, \varsigma_{n,k}^{(\Phi)})$ using $\mathbf{P}_{\text{FAP}}^{(\Phi)}$, NOMA decoding order.
        \STATE \quad \quad Solve convex approximation of \eqref{powerequtran} for $\mathbf{P}_{\text{FAP}}^{(\Phi+1)}$.
        \STATE \quad \quad Check convergence with tolerance $\epsilon_{\text{FAP}}$. Set $\Phi = \Phi + 1$.
        \STATE \quad \textbf{End While}
        \STATE \textbf{CP Power Allocation:}
        \STATE \quad \quad Set $\Phi = 0$.
        \STATE \quad \quad \textbf{While} $\Phi < \Phi_{\max}$ and not converged \textbf{do}
        \STATE \quad \quad \quad Update $(\upsilon_{\text{CP},k}^{(\Phi)}, \varsigma_{\text{CP},k}^{(\Phi)})$ using $\mathbf{P}_{\text{CP}}^{(\Phi)}$.
        \STATE \quad \quad \quad Solve convex approximation of \eqref{4_powereCP} for $\mathbf{P}_{\text{CP}}^{(\Phi+1)}$.
        \STATE \quad \quad \quad Check convergence with tolerance $\epsilon_{\text{CP}}$. Set $\Phi = \Phi + 1$.
        \STATE \quad \quad \textbf{End While}
        \STATE \textbf{Output:} Optimized powers $\mathbf{P}_{\text{FAP}}^*$ and $\mathbf{P}_{\text{CP}}^*$.
    \end{algorithmic}
\end{algorithm}
	
The computational complexity of the proposed SCA-based power allocation algorithm, outlined in Algorithm \ref{4_IRLS}, is driven by the convex optimization steps for FAP and CP power allocations. The algorithm optimizes power for \( K \) users, \( N \) FAPs, and \( F \) content files across \( N+1 \) nodes (CP and FAPs). Using CVX with an interior-point method, the per-iteration complexity includes: Optimizes \( P_{\text{FAP}} \in \mathbb{R}^{N \times K} \) for users served by \( N \) FAPs. Assuming \( K/N \) users per FAP, the complexity per FAP is \( \mathcal{O}((K/N)^{3.5}) \), yielding a total of \( \mathcal{O}(K^{3.5} / N^{2.5}) \) across \( N \) FAPs. Optimizes \( P_{\text{CP}}.\text{users} \) for \( 1 \) CP users and \( P_{\text{CP}}.\text{files} \) for \( F \) files, with complexity \( \mathcal{O}((1 + F)^{3.5}) \). The per-iteration complexity, combining sequential FAP and CP optimizations, is:
\(\mathcal{O} ( K^{3.5} / N^{2.5} + \left( 1 + F \right)^{3.5} ).\)
With a maximum of \( \Phi_{\text{max}} \) iterations, the total complexity is:
\(\mathcal{O} ( \Phi_{\text{max}}  ( K^{3.5} / N^{2.5} + \left( 1 + F \right)^{3.5} ) ). \) Compared to an exhaustive search (\( \mathcal{O}(2^{N K + K + F}) \)), the proposed algorithm’s polynomial complexity ensures scalability.

\subsection{Overall Algorithm}

In this subsection, we present a comprehensive overview of the proposed AO algorithm that integrates user association, cache placement, and power allocation to minimize average delay. The algorithm iteratively solves subproblems to achieve a near-optimal solution, balancing computational complexity and performance. 

The iterative optimization process is outlined as follows: 1) \textbf{Initialization:} Begin with an initial configuration for user association, cache placement, and power allocation. 2) \textbf{User Association and Cache Placement Optimization:} Employ Algorithm \ref{4_JACPM} to optimize user association and cache placement based on the current power allocation coefficients. 3) \textbf{Power Allocation Optimization:} Utilize Algorithm \ref{4_IRLS} to refine power allocation coefficients, given the fixed user association and cache placement. This step leverages SCA to iteratively improve power allocation. 4) \textbf{Iteration:} Alternate between Algorithm \ref{4_JACPM} and Algorithm \ref{4_IRLS}, updating user association, cache placement, and power allocation in each iteration.

The total complexity depends on the inner iterations of the subproblems and the outer iterations of the AO algorithm. The AO algorithm alternates between Algorithm \ref{4_JACPM} and Algorithm \ref{4_IRLS}. Let \( T_{\text{outer}} \) denote the number of outer iterations. Thus, the total complexity of the AO algorithm is: 
$\mathcal{O} ( T_{\text{outer}}  ( \frac{1}{\varepsilon}  \max ( K^3, N F^{3.5}, N K F )  +  \Phi_{\text{max}}  ( K^{3.5} / N^{2.5} + \left( 1 + F \right)^{3.5} ) ) )$. Although the complexity expression involves multiple terms, it is dominated by low-order polynomial terms $\mathcal{O}(K^{3.5} + N F^{3.5})$ per iteration in practical settings where $K$, $N$, and $F$ are moderate (e.g., tens of users and files, as typical in edge networks). This complexity scaling aligns with standard interior-point methods for convex optimization \cite{boyd2004convex}. This is fundamentally different from exponential-complexity global solvers (e.g., branch-and-bound used in LINGO) or NP-hard exhaustive search with complexity $\mathcal{O}(2^{N K F} \cdot K!)$, which are verified to be computationally prohibitive even for small-scale instances due to the combinatorial nature of MINLP \cite{wu2022non}. Thus, the proposed JACPM+SCA framework achieves an excellent trade-off between near-optimal performance and practical scalability, fully justifying its classification as a low-complexity solution suitable for real-time deployment in 6G edge networks.

\section{Numerical Results and Discussions}\label{VI}
This section presents the simulation results and validates the effectiveness of the proposed algorithm. In the simulation scenario setup, a circular area with a radius of $R_D=500m$ is considered, with the CP located at the center of the circle. The positions of the three fixed FAPs are at $(0, 0.5R_D)$, $(0.4R_D, -0.3R_D)$, and $(-0.4R_D, -0.3R_D)$. The UEs are uniformly and independently distributed, with a total number of $K=7$. Users request content files based on the Zipf probability distribution. Considering the file popularity model, the request probability for the $i$-th file is given by $P(F_i)=\frac{\frac{1}{i^r}}{\sum_{p=1}^{F}\frac{1}{p^r}}$, where $r>0$ represents the skewness of content popularity; the larger the Zipf parameter, the more uneven the popularity distribution. Each simulation result is averaged over 100 channel realizations. The transmit power, noise power, path loss model, and other physical layer parameters adhere to 3GPP standards and are informed by cutting-edge research in F-RANs~\cite{qiu2021subchannel}. Unless otherwise specified, other parameters used in the simulation are summarized in Table \ref{4_T}.\footnote{{
In this simulation section, we focus on quantifying the performance gains achieved specifically by the proposed resource allocation algorithm. Since $D_{\text{ov}}$ is a constant value determined by protocol standards and cannot be reduced through the optimization of decision variables, and considering it is generally negligible compared to the transmission delay in the examined scenarios, we assume $D_{\text{ov}} \approx 0$ in the following numerical results.}}
\begin{table}[!t]
	\vspace{-0.2cm}\small
	\renewcommand{\arraystretch}{1.3}
	\caption{System Simulation Parameters}
	\label{4_T}
	\begin{center}\vspace{-0.4cm}
	\begin{tabular}{|p{4.2cm}<{\centering}|p{3.7cm}<{\centering}|}
	\hline
	\textbf{Parameter} & \textbf{Value} \\ \hline
	{Transmit power of the CP} & $40$~dBm \\\hline
	{Transmit power of the FAP} & $30$~dBm \\\hline
	Path loss model for the CP & $15.3+37.6\log_{10}(d),d(m)$ \\\hline
	Path loss model for the FAP & $38.46+20\log_{10}(d),d(m)$ \\\hline
	{Noise power spectral density} $\sigma^2$ & $-174$~dBm/Hz \\\hline
	Bandwidth & $10$~MHz \\\hline
	File size & $10$~Kb \\\hline
	Cache capacity $S_n$ & $20$~Kb\\\hline
	User capacity $F_n$ & $(1,2,2,2)$\\\hline
	Number of files & $10$ \\\hline
	Number of users & $7$\\\hline
	Zipf skewness (\( \gamma \)) & 0.8 \\ \hline
	\end{tabular}
	\end{center}
	\vspace{-0.2cm}
\end{table}
In the proposed NOMA-based F-RANs, we evaluate its performance against three benchmark schemes: OMA-based F-RANs, fixed-power NOMA-based F-RANs and Most Popular Content Maximum SINR (MCP-MS) scheme.
\begin{itemize}
\item \textbf{OMA}: In this scheme, TDMA is employed as the multiple access technique at the access points. The time slots for the UEs' received signals are evenly distributed, with each UE assigned a time slot allocation factor of $1/K$.
\item \textbf{Fixed-NOMA}: For any cluster $\mathcal{K}_n$ containing $t$ UEs in a cell, the power allocation coefficient for the $k$-th UE is defined as $p_{n,i}=\frac{t-i+1}{\mu}$, where $\mu=\sum_{i=1}^{t}i$ ensures $\sum_{i=1}^{t}p_{n,i}=1$.
\item \textbf{MCP-MS}: The benchmark scheme MCP-MS\cite{yang2015analysis}, refers to users selecting the access node based on the maximum SINR criterion and caching the most popular file content.
\end{itemize}

The effectiveness of the proposed algorithm, combining joint user association and cache placement (JACPM) with successive convex approximation (SCA) power allocation (JACPM + SCA), will be evaluated by comparing its performance against these benchmark schemes.
\begin{figure}[!htb]
	\centering  
	\subfigure[Cache capacity = 0 Kb]{  
	\begin{minipage}{0.45\columnwidth}
	\centering    
	\includegraphics[width=\linewidth]{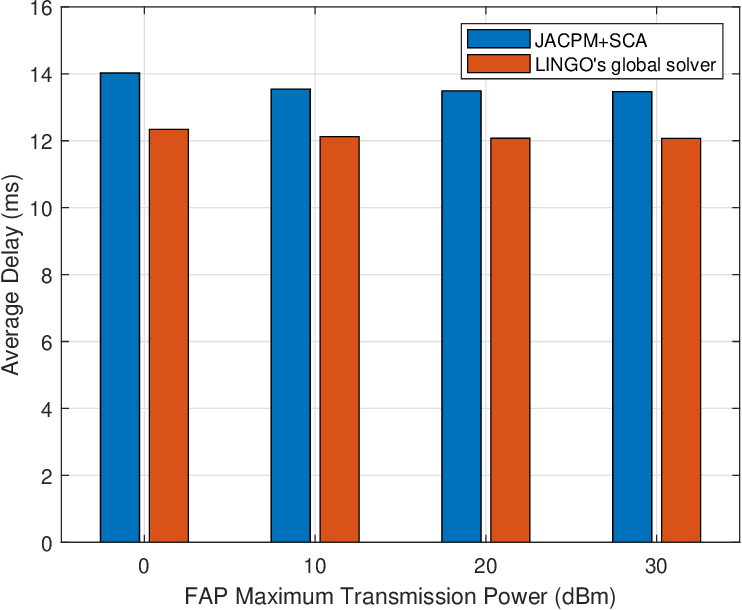}  
	\end{minipage}}
	\hfill
	\subfigure[Cache capacity = 20 Kb]{ 
	\begin{minipage}{0.45\columnwidth}
	\centering   
	\includegraphics[width=\linewidth]{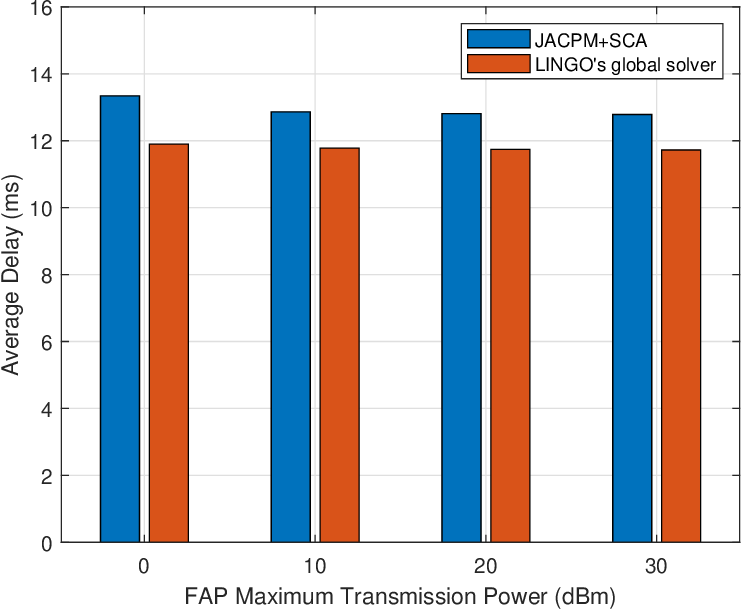}
	\end{minipage}}
	\caption{Average delay under different maximum transmission power of FAP}   
	\label{fig:delaylingo}    
\end{figure}

Fig. \ref{fig:delaylingo} evaluates the average transmission delay performance of the proposed JACPM+SCA algorithm against the LINGO global solver under varying FAP maximum transmit powers. The evaluation considers two cache capacity scenarios: 0 Kb and 20 Kb. The LINGO solver serves as a benchmark\cite{wu2022non}, employing an general branch-and-bound approach to provide a baseline for delay performance, at a significantly higher computational cost. The proposed JACPM+SCA algorithm consistently reduces the average delay across all tested transmit power levels (0 dBm to 30 dBm). Although LINGO yields a slightly lower delay due to its global optimization nature, the marginal gap demonstrates that JACPM+SCA effectively approximates the optimal solution with substantially lower computational overhead. Fig. \ref{fig:delaylingo}(b) presents the delay performance for a cache capacity of 20 Kb. Here, JACPM+SCA records an average delay of 12.7835 ms at 30 dBm, compared to LINGO's 11.7235 ms, a difference of 1.0600 ms (approximately 9.0\% higher). The smaller performance gap with increased cache capacity highlights JACPM+SCA's ability to leverage caching effectively, further reducing transmission delays and approaching the globally optimal solution provided by LINGO.

\begin{table}[!htb]
    \vspace{-0.2cm}\small
    \renewcommand{\arraystretch}{1.3}
    \caption{Computation Time Comparison at FAP Maximum Power of 30 dBm}
    \label{tab:time_comparison_30dBm}
    \begin{center}\vspace{-0.4cm}
    \begin{tabular}{|p{2.8cm}<{\centering}|p{2.5cm}<{\centering}|p{2cm}<{\centering}|}
        \hline
        Cache Capacity (Kb) & JACPM+SCA (s) & LINGO (s) \\
        \hline
        0 & 12.89 & 7162\\
        \hline
        20 & 3.83 & 1898 \\
        \hline
    \end{tabular}
    \end{center}
    \vspace{-0.2cm}
\end{table}

The computational efficiency of JACPM+SCA is further substantiated in Table \ref{tab:time_comparison_30dBm}, which compares computation times at FAP transmit power of 30 dBm. For a cache capacity of 0 Kb, JACPM+SCA requires only 12.89 s, while LINGO demands 7162 s (approximately 2 hours), achieving a computational speedup exceeding 550 times. With a cache capacity of 20 Kb, JACPM+SCA completes in 3.83 s, compared to LINGO's 1898 s, yielding a speedup of nearly 500 times. This dramatic reduction in computation time positions JACPM+SCA as a highly practical solution for real-time applications in large-scale wireless networks.
\begin{figure}[!htb]
	\centering
	\includegraphics[width=0.35\textwidth]{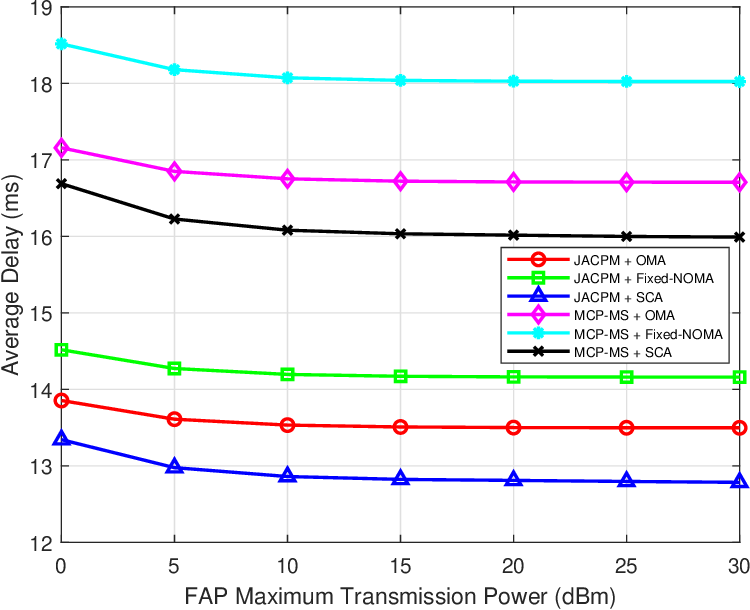}\\
	\caption{Average transmission delay versus FAP maximum transmission power for different schemes.}\label{delaypoweroma}
\end{figure}

Fig. \ref{delaypoweroma} evaluates the average transmission delay performance of different schemes under varying FAP maximum transmission powers, ranging from 0 dBm to 30 dBm. The proposed JACPM+SCA consistently outperforms all benchmark schemes across the tested power range. Specifically, at a FAP maximum transmit power of 30 dBm, the proposed scheme achieves an average delay of 12.7835 ms, compared to 13.4967 ms for JACPM + OMA, 14.1593 ms for JACPM + Fixed-NOMA, and 15.9902 ms for MCP-MS + SCA. This corresponds to delay reductions of approximately 5.3\%, 9.7\%, and 20.1\%, respectively. The performance comparison with the OMA benchmark acts as an ablation study, clearly quantifying the benefits brought by NOMA. The achieved latency reduction confirms that power-domain multiplexing successfully enables concurrent content pushing and user request servicing, thereby overcoming the resource partitioning overhead of OMA and markedly improving system-wide efficiency. Furthermore, the analysis reveals that the transmission delay decreases noticeably as the FAP transmission power increases within the low-power region. For instance, the delay for JACPM + SCA drops from 13.3416 ms at 0 dBm to 12.8598 ms at 10 dBm, a reduction of approximately 3.6\%. However, this improvement diminishes in the high-power region, where the delay decreases from 12.8090 ms to 12.7835 ms, further delay reduction becomes marginal due to the interference-limited nature of NOMA clusters. When the FAP transmit power exceed approximately 10~dBm, as the transmit power increases, both the desired signal and the intra-cluster/inter-cluster interference increase proportionally, resulting in an almost constant SINR and thus a saturation of the achievable rate and delay.

\begin{figure}[!htb]
	\centering
	\includegraphics[width=0.35\textwidth]{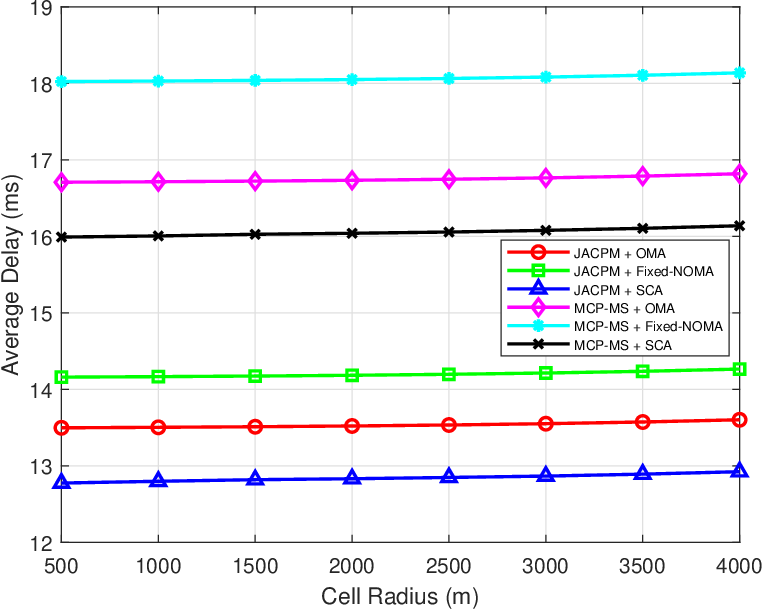}\\
	\caption{Average transmission delay versus cell radius for different schemes.}\label{delayradius}
\end{figure}

In the following, we conduct a comprehensive sensitivity analysis to evaluate how the system performance responds to variations in key physical and network parameters. This analysis validates the robustness of the proposed model and algorithm under diverse network conditions. Fig. \ref{delayradius} analyzes the impact of cell radius on average transmission delay. The simulation evaluates the performance of the proposed JACPM + SCA scheme against several benchmarks. The cell radius varies from 500 m to 4000 m. As the cell radius increases, the average channel gain diminishes due to heightened path loss, leading to reduced transmission rates and elevated delays across all schemes. At a radius of 4000 m, JACPM + SCA records a delay of 12.9229 ms, compared to 13.6018 ms for JACPM + OMA (a 5.0\% relative increase) and 16.1382 ms for MCP-MS + SCA (a 24.8\% relative increase). The widening performance gap highlights the adaptability of JACPM + SCA to worsening channel conditions. As radius decreases, the rise in signal power is largely offset by a corresponding increase in interference power, stabilizing the SINR across the range of cell radii. Since the transmission rate depends logarithmically on SINR and delay is inversely related to the rate, this stability translates to a gradual rather than pronounced change in delay. This trend is consistent across all schemes in Fig. \ref{delayradius}, where the shallow slopes of delay curves reflect the counterbalancing effects of channel gain and interference.
\begin{figure}[!htb]
	\centering
	\includegraphics[width=0.35\textwidth]{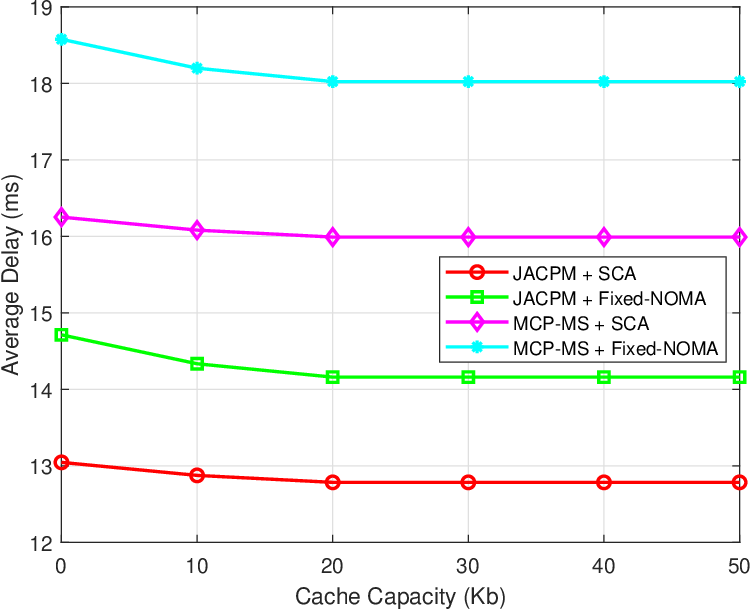}\\
	\caption{Average transmission delay versus FAP cache capacity for different schemes.}\label{delaycache}
\end{figure}

Fig. \ref{delaycache} illustrates the relationship between average transmission delay and FAP cache capacity. The proposed JACPM + SCA scheme consistently achieves the lowest transmission delay across all cache capacities. At a cache capacity of 20 Kb, JACPM + SCA records an average delay of 12.7835 ms, reducing delay by 29.2\% compared to MCP-MS + Fixed-NOMA (18.0235 ms), 20.1\% compared to MCP-MS + SCA (15.9902 ms), and 9.7\% compared to JACPM + Fixed-NOMA (14.1593 ms). Beyond 20 Kb, the delay for all schemes stabilizes, indicating that the cache capacity is sufficient to store all requested content, eliminating additional fronthaul delays. The decrease in delay with increasing cache capacity is attributed to the expanded feasible region for content caching. Larger cache capacities allow FAPs to store more files, increasing the likelihood of satisfying user requests locally and reducing reliance on fronthaul links, which incur higher delays. 
\begin{figure}[!htb]
	\centering
	\includegraphics[width=0.35\textwidth]{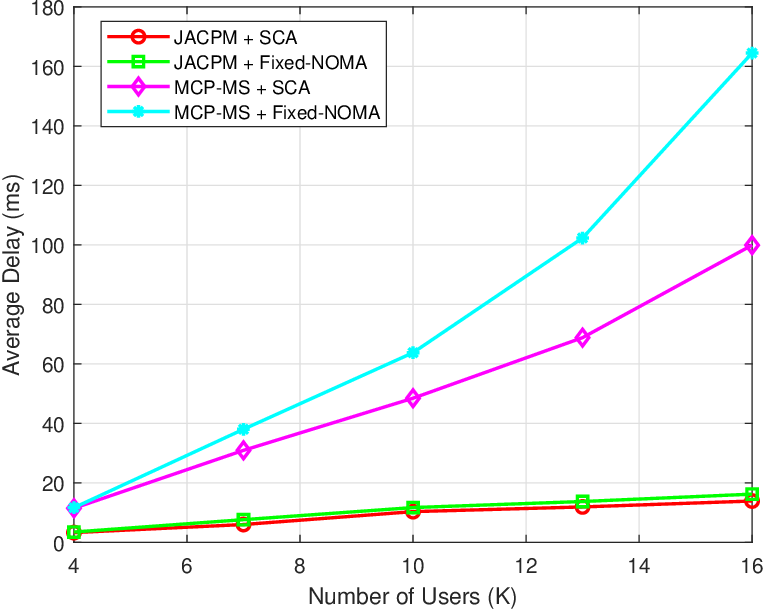}\\
	\caption{Average transmission delay versus number of users for different schemes.}\label{delaynumber}
\end{figure}

Fig. \ref{delaynumber} examines the impact of the number of users \( K \) on the average transmission delay. As shown in Fig. \ref{delaynumber}, the proposed JACPM + SCA scheme consistently achieves the lowest transmission delay across all user counts. Notably, JACPM + SCA reduces delay by 14.3\% compared to JACPM + Fixed-NOMA, 86.1\% compared to MCP-MS + SCA, and 91.6\% compared to MCP-MS + Fixed-NOMA at \( K = 16 \), highlighting its superior scalability. The increase in delay with more users is driven by heightened content request demands, which strain the limited FAP cache capacity. As \( K \) grows, the diversity of file requests increases, reducing cache hit rates and necessitating more fronthaul transmissions, which incur higher delays. Additionally, in NOMA-based F-RANs, the increased number of users per cluster amplifies intra-cluster interference, further reducing transmission rates. 
\begin{figure}[!htb]
	\centering
	\includegraphics[width=0.35\textwidth]{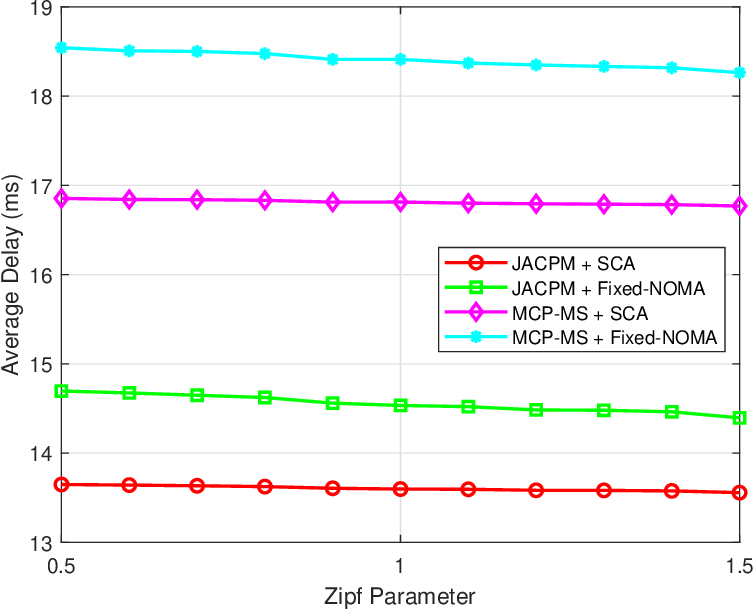}\\
	\caption{Average transmission delay versus Zipf parameter for different schemes.}\label{delayzipf}
\end{figure}

Fig. \ref{delayzipf} evaluates the impact of content popularity skewness, characterized by the Zipf parameter, on average transmission delay. The average transmission delay decreases as the Zipf parameter increases, reflecting a more skewed content popularity distribution. At a Zipf parameter of 0.9, JACPM + SCA achieves a delay of 13.6052 ms, reducing delay by 25.8\% compared to MCP-MS + Fixed-NOMA (18.4117 ms), 19.1\% compared to MCP-MS + SCA (16.8120 ms), and 6.6\% compared to JACPM + Fixed-NOMA (14.5593 ms). The reduction in delay with increasing Zipf parameter is attributed to enhanced caching efficiency. A higher Zipf parameter concentrates user requests on a smaller subset of popular files, increasing the likelihood that requested content is cached at FAPs. This reduces reliance on fronthaul links, which incur higher delays.

\section{Conclusion}\label{VII}
In this work, a novel low-delay service framework for NOMA-based F-RANs was proposed, leveraging the capabilities of NOMA within a fog-cloud architecture. The framework was designed to address the complex challenges associated with joint resource allocation, including user association, cache placement, and power allocation. By employing an alternating optimization algorithm, which decomposed the original problem into two subproblems, we effectively tackled the non-convex MINLP problem. The proposed framework demonstrated significant improvements over conventional OMA-based F-RAN systems, particularly in reducing average transmission delay across various scenarios. The insights gained from our simulations underscored the importance of a low-complexity approach to optimize user association and cache placement, as well as the effectiveness of SCA in power allocation. Moreover, the proposed algorithm achieved an optimal balance between performance and computational efficiency, highlighting its potential for practical deployment in next-generation wireless networks. 

\appendices

\bibliographystyle{IEEEtran}
\bibliography{IEEEabrv, egbib}
\begin{IEEEbiography}[{\includegraphics[width=1in,height=1.25in,clip,keepaspectratio]{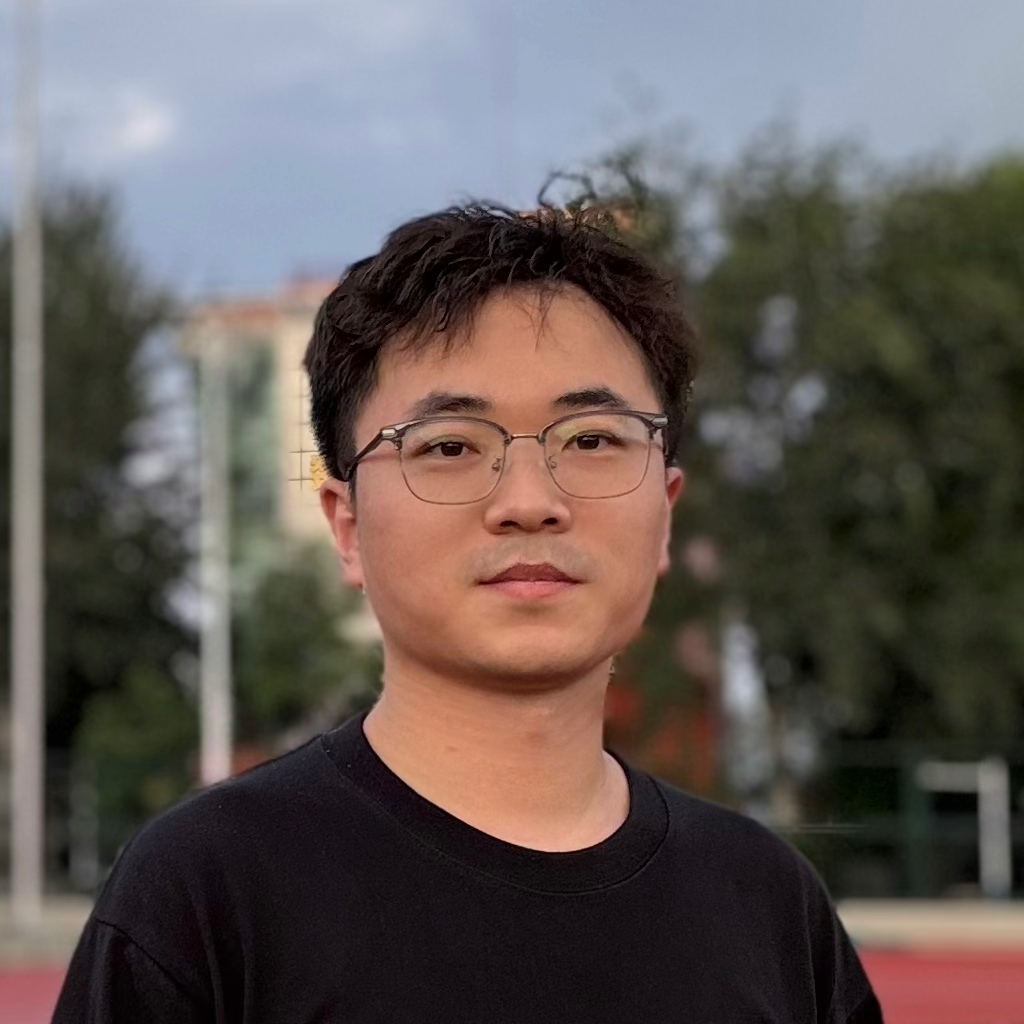}}]
    {Yuan Ai} received the Ph.D. degree in information and communication engineering from the Beijing University of Posts and Telecommunications (BUPT), Beijing, China, in 2023. From 2023 to 2025, he was a Postdoctoral Researcher with the School of Information Science and Technology, Beijing University of Technology (BJUT), Beijing. From 2024 to 2025, he was also a Visiting Research Associate with the Department of Electrical and Electronic Engineering, The University of Hong Kong (HKU). He is currently a Lecturer with the School of Information Science and Technology, BJUT. His research interests include flexible-antenna technologies, next-generation multiple access (NGMA), AI for B5G/6G, and edge intelligence.
\end{IEEEbiography}  
\begin{IEEEbiography}[{\includegraphics[width=1in,height=1.25in,clip,keepaspectratio]{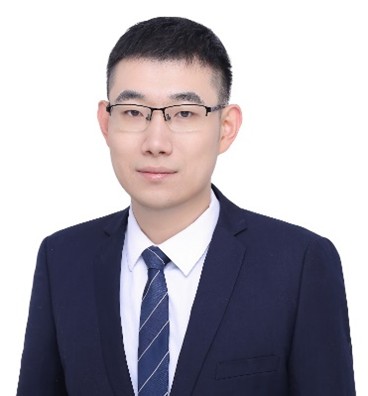}}]
    {Xidong Mu} (Member, IEEE, \url{https://xidongmu.github.io/}) received the Ph.D. degree in Information and Communication Engineering from the Beijing University of Posts and Telecommunications, Beijing, China, in 2022. He was with the School of Electronic Engineering and Computer Science, Queen Mary University of London, from 2022 to 2024, where he was a Postdoctoral Researcher. He has been a lecturer (an assistant professor) with the Centre for Wireless Innovation (CWI), School of Electronics, Electrical Engineering and Computer Science, Queen’s University Belfast, U.K. since August 2024. His research interests include flexible-antenna technologies, reconfigurable surface aided communications, next generation multiple access (NGMA), integrated sensing and communications, and optimization theory. 

    Xidong Mu is a Web of Science Highly Cited Researcher. He received the IEEE ComSoc Outstanding Young Researcher Award 
    for EMEA region in 2023 and the IEEE ComSoc Wireless Communications Technical Committee (WTC) Outstanding Young Researcher Award in 2025. He is the recipient of the 2024 IEEE Communications Society Heinrich Hertz Award, the Best Paper Award in ISWCS 2022, the 2022 IEEE SPCC-TC Best Paper Award, and the Best Student Paper Award in IEEE VTC2022-Fall. 
    He serves as the secretary of the IEEE ComSoc Technical Committee on Cognitive Networks (TCCN), the secretary of the IEEE ComSoc NGMA Emerging Technology Initiative, and the URSI UK Early Career Representative (ECR) for Commission C. He also serves as an Editor of \textsc{IEEE Transactions on Communications}, a Guest Editor for \textsc{IEEE Journal on Selected Areas in Communications}, \textsc{IEEE Transactions on Cognitive Communications and Networking}, \textsc{IEEE Internet of Things Journal}, and the ``Mobile and Wireless Networks'' symposium co-chair of IEEE GLOBECOM 2025.     
\end{IEEEbiography}
\begin{IEEEbiography}[{\includegraphics[width=1in,height=1.25in,clip,keepaspectratio]{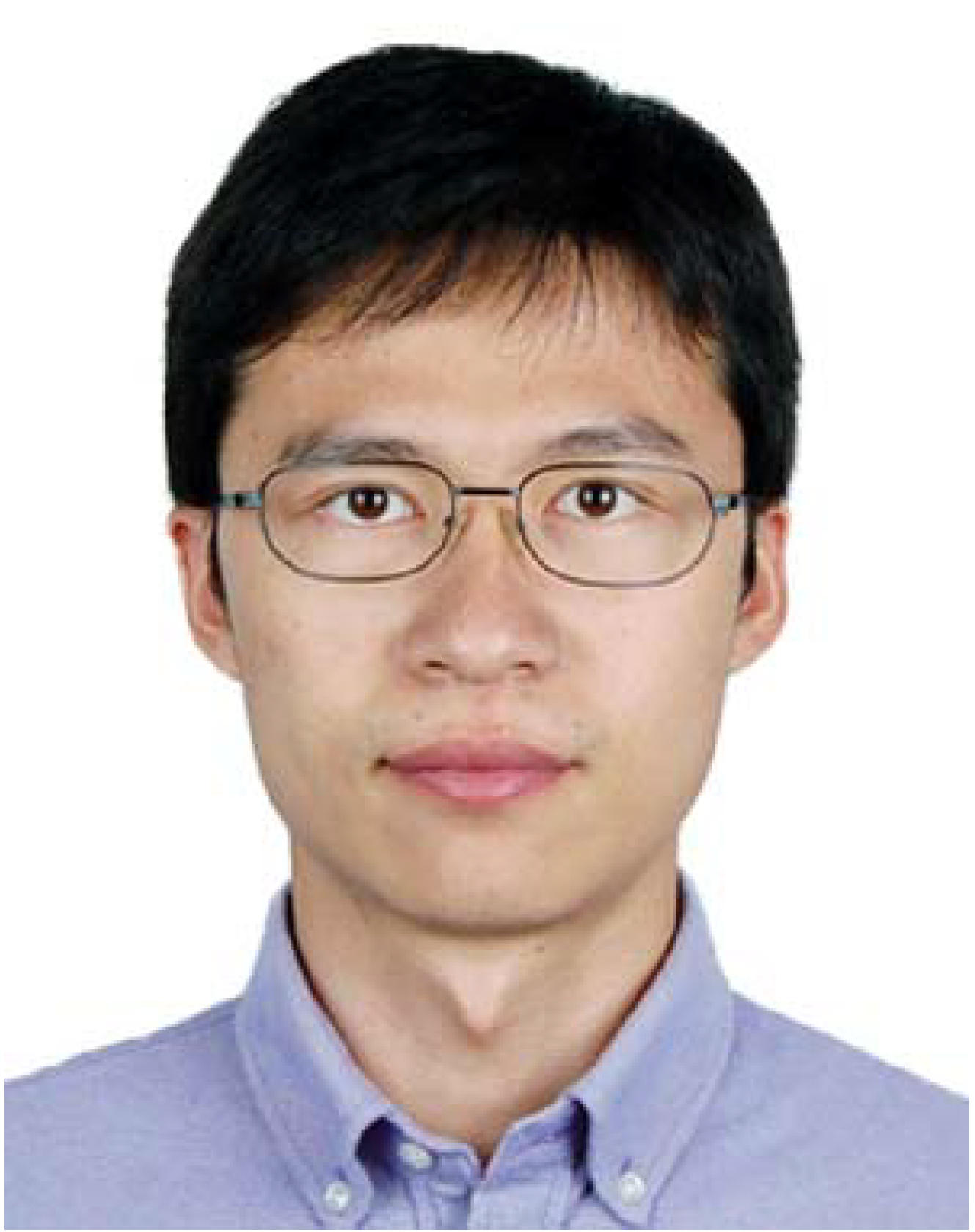}}]
{Pengbo Si} (Senior Member, IEEE) received the B.S. and Ph.D. degrees from the Beijing University of Posts and Telecommunications (BUPT), Beijing, China, in 2004 and 2009, respectively. He joined the Beijing University of Technology (BJUT), Beijing, in 2009, where he is currently a Professor. From 2007 to 2008, he was a Visiting Student with Carleton University, Ottawa, ON, Canada. From 2014 to 2015, he was a Visiting Scholar with the University of Florida, Gainesville, FL, USA. His research interests include blockchain, SDN, resource management, and cognitive radio networks. Dr. Si serves as an Associate Editor for the \textit{International Journal on Ad Hoc Networking Systems} and an Editorial Board Member for \textit{Ad Hoc \& Sensor Wireless Networks}. He was a Symposium Chair for IEEE GLOBECOM 2019. He has served as a Guest Editor for \textit{Advances in Mobile Cloud Computing} and a Special Issue of \textit{IEEE Transactions on Emerging Topics in Computing}. He was also the TPC Co-Chair of IEEE ICCC'13-GMCN, the Program Vice Chair of IEEE GreenCom'13, and a TPC Member for numerous international conferences.
\end{IEEEbiography}  
\begin{IEEEbiography}[{\includegraphics[width=1in,height=1.25in,clip,keepaspectratio]{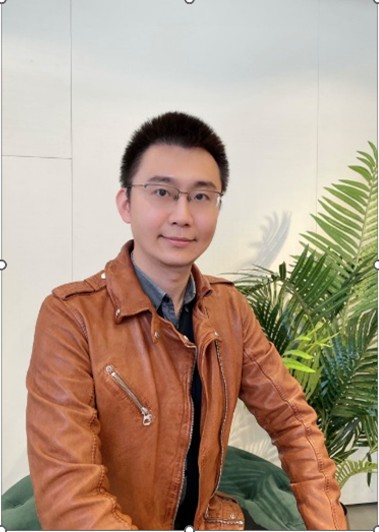}}]
     {Yuanwei Liu} (S'13-M'16-SM'19-F'24, \url{https://www.eee.hku.hk/~yuanwei/}) is a tenured full Professor in Department of Electrical and Electronic Engineering (EEE) at 
     The University of Hong Kong (HKU) and a visiting professor at Queen Mary University of London (QMUL). 
     Prior to that, he was a Senior Lecturer (Associate Professor) (2021-2024) and a Lecturer (Assistant Professor) (2017- 2021) at QMUL, London, U.K, 
     and a Postdoctoral Research Fellow (2016-2017) at King's College London (KCL), London, U.K. He received the Ph.D. degree from QMUL in 2016.  
     His research interests include non-orthogonal multiple access, reconfigurable intelligent surface, near field communications, integrated sensing and communications, and machine learning. 
 
     Yuanwei Liu is a Fellow of the IEEE, a Fellow of AAIA, a Fellow of AIIA, a Web of Science Highly Cited Researcher, an IEEE Communication Society Distinguished Lecturer, 
     an IEEE Vehicular Technology Society Distinguished Lecturer, the rapporteur of ETSI Industry Specification Group on Reconfigurable Intelligent Surfaces on work item of ``Multi-functional Reconfigurable Intelligent Surfaces (RIS): Modelling, Optimisation, and Operation'', 
     and the UK representative for the URSI Commission C on ``Radio communication Systems and Signal Processing'' (2023-2024). He was listed as one of 35 Innovators Under 35 China in 2022 by MIT Technology Review. He received IEEE ComSoc Outstanding Young Researcher Award for EMEA in 2020. He received the 2020 IEEE Signal Processing and Computing for Communications (SPCC) Technical Committee Early Achievement Award, IEEE Communication Theory Technical Committee (CTTC) 2021 Early Achievement Award. He received IEEE ComSoc Outstanding Nominee for Best Young Professionals Award in 2021. He is the co-recipient of the 2024 IEEE Communications Society Heinrich Hertz Award, the Best Student Paper Award in IEEE VTC2022-Fall, the Best Paper Award in ISWCS 2022, the 2022 IEEE SPCC-TC Best Paper Award, the 2023 IEEE ICCT Best Paper Award, and the 2023 IEEE ISAP Best Emerging Technologies Paper Award. He serves as the Co-Editor-in-Chief of IEEE ComSoc TC Newsletter, an Area Editor of IEEE Transactions on Communications and IEEE Communications Letters, an Editor of IEEE Communications Surveys \& Tutorials, IEEE Transactions on Wireless Communications, IEEE Transactions on Vehicular Technology, IEEE Transactions on Network Science and Engineering, and IEEE Transactions on Cognitive Communications and Networking. He serves as the (leading) Guest Editor for Proceedings of the IEEE on Next Generation Multiple Access, IEEE JSAC on Next Generation Multiple Access, IEEE JSTSP on Intelligent Signal Processing and Learning for Next Generation Multiple Access, and IEEE Network on Next Generation Multiple Access for 6G. He serves as the Publicity Co-Chair for IEEE VTC 2019-Fall, the Panel Co-Chair for IEEE WCNC 2024, Symposium Co-Chair for several flagship conferences such as IEEE GLOBECOM, ICC and VTC. He serves the academic Chair for the Next Generation Multiple Access Emerging Technology Initiative, vice chair of SPCC and Technical Committee on Cognitive Networks (TCCN) (2023-2024).
\end{IEEEbiography}
\end{document}